\documentclass[preprint]{elsarticle}
\usepackage{graphicx}
\usepackage{epsf}
\usepackage{wrapfig}
\usepackage{epsfig}

\def\b0{{\mbox{\boldmath$0$}}}

\usepackage{color}

\usepackage{ulem}

\usepackage{graphicx}
\usepackage{epsf}
\usepackage{epstopdf}
\usepackage{wrapfig}
\usepackage{epsfig}
\usepackage{epsfig}
\def\Vec#1{\mbox{\boldmath $#1$}}

\def\beq{\begin{equation}}
\def\eeq{\end{equation}}

\def\beqy{\begin{eqnarray}}
\def\eeqy{\end{eqnarray}}

\newcommand{\slfrac}[2]{\left.#1\middle/#2\right.}

\def \b #1{ {\bf #1}}
\newcommand{\be}{\begin{eqnarray}}
\newcommand{\ee}{\end{eqnarray}}

\def \b #1{ {\bf #1}}

\def \b #1{ {\bf #1}}
     \font\tenbifull=cmmib10 scaled 1200 % bold math italic
     \font\tenbimed=cmmib9
     \font\tenbismall=cmmib7
\mathchardef\bbkappa="7114
\mathchardef\bbrho="711A
\mathchardef\bbsigma="711B
\mathchardef\bbtau="711C
\mathchardef\bbvarrho="7125
\mathchardef\bbvarsigma="7126
\mathchardef\bbxi="7118

\begin{document}
\vskip 2mm \date{\today}\vskip 2mm
\title{Color fluctuation effects in proton-nucleus collisions}
\author[ect,irpi]{M. Alvioli}
\address[ect]{ECT$^\star$, European Centre for Theoretical Studies in Nuclear Physics 
  and Related Areas, Strada delle Tabarelle 286, I-38123 Villazzano (TN) Italy}
\address[irpi]{CNR IRPI, via Madonna Alta 126, 06128 Perugia, Italy}
\author[psu]{M. Strikman}
\address[psu]{104 Davey Lab, The Pennsylvania State University,
  University Park, PA 16803, USA}
\vskip 2mm
%------------------------------------------------------------------------------%
%------------------------------------------------------------------------------%
\begin{abstract}
Color fluctuations in hadron-hadron collisions are responsible for the presence of inelastic diffraction and lead 
to distinctive differences  between  the Gribov picture of high energy scattering and  the low energy Glauber picture. 
We find that color fluctuations give a larger  contribution to the fluctuations of the number of wounded nucleons
than the fluctuations of the number of nucleons at a given impact parameter. The two contributions for the impact 
parameter averaged fluctuations are comparable. As a result, standard procedures for selecting 
peripheral (central) collisions lead to selection of configurations in the projectile which interact with smaller 
(larger) than average strength. We suggest that studies of $pA$ collisions with a hard trigger may allow to observe 
effects of color fluctuations.
\end{abstract}
\maketitle
\section{Introduction}\label{sec:sec1}
Currently most of the experimental studies as well as modeling of the nucleus-nucleus (proton-nucleus) collisions 
involve using the Glauber model. Namely, the number of involved nucleons is calculated probabilistically assuming 
that each Nucleon-Nucleon (NN) inelastic collision is determined by the value of $\sigma_{in}^{NN}$
at the collision energy.

However, the dominance of large longitudinal distances in high energy scattering  \cite{GPI} changes qualitatively the 
pattern of multiple interactions. Indeed, in the Glauber approximation high energy interactions of the projectile with 
a target occur via consecutive rescatterings of the projectile off the constituents of the target. The projectile during
the interactions is on mass shell -- one takes the residues in the propagators of the projectile.
This approximation contradicts the QCD based space-time evolution of high 
energy processes dominated by particles production. The projectile interacts with the target in  frozen configurations 
since the life time of the configurations becomes much larger than the size of the target. Hence there is no time for 
a frozen configuration in the projectile to combine back into the projectile during the time of the order $R_T$, the 
radius of the target. 
As a result the amplitudes described by Glauber model diagrams die out at large energies  $\propto 1/s$ (a formal proof which 
is based on the analytic properties of the Feynman diagrams was given in \cite{S.Mandelstam63,PomeronCalculus}).  

In the Glauber model the number of interacting  nucleons is calculated probabilistically assuming that the probability 
of individual NN inelastic collisions is determined by the value of $\sigma_{in}^{NN}$
at the collision energy. Fluctuations of the number of wounded nucleons at a given impact parameter are generated 
solely by fluctuations of the positions of nucleons in the nucleus and (in some models) due to peripheral 
collisions of nucleons, where the interaction is gray and hence the chance to interact differs from one or zero. 
Hard collisions are treated as  binary collisions, which is equivalent to taking the diagonal generalized 
parton densities of nuclei, $f_A(x,Q^2, b)$, proportional to the impact factor $T(b)$: 
\beq
F_A(x,Q^2, b)= f_N(x,Q^2)\,T(b),
\eeq
where $T(b)$ is normalized as $\int d\Vec{b}\,T(b)=A$. A nuclear shadowing correction is introduced for $x\le 0.01$.

The high energy theory of soft interactions with nuclei was developed by Gribov \cite{Gribov:1968jf} who expressed 
the shadowing contribution to the cross section of hadron-nucleus (hA) interactions through the contribution of non-planar 
diagrams. The Gribov-Glauber theory, in difference from the low energy Glauber theory, requires taking into account that a
particular quark-gluon configuration of the projectile is frozen during the collision and that  it may interact with 
different strength as compared to the average strength. This leads to fluctuations of the number of collisions which 
are significantly larger than in the Glauber model. The fluctuations of the strength of the interaction are related 
to the ratio of inelastic and elastic diffraction in NN scattering  at $t=0$. Relevance of fluctuations of the 
strength was first pointed out in \cite{Baym:1994uh,Baym:1995cz} but these effects were never analyzed in detail before. 

Another effect contributing to fluctuations of observables in hA collisions is fluctuations of the gluon density which 
can originate both from the fluctuations of the nucleon configurations and from the fluctuations of the gluon densities in 
the individual nucleons. We will consider this effect elsewhere.

The paper is organized as follows. In section \ref{sec:sec2} we summarize the necessary information about fluctuations 
of the strength of NN interaction. In section \ref{sec:sec3} we use the Gribov-Glauber model in the optical approximation 
to obtain analytic results for the strength of fluctuations of the number of wounded nucleons and relative contributions 
to these fluctuations of the fluctuations of the strength of the interaction and of geometry of collisions.
In section \ref{sec:sec4} we develop a full Monte Carlo (MC) model in which the geometry of projectile--target 
nucleon interaction is accounted for, and the strength of the interaction fluctuates on an event-by-event basis.  
The results for the change of the distribution over the number of collisions and for the dependence of the average 
strength of the interaction on impact parameter are presented. Possibilities for observing color fluctuation 
effects in collisions with hard triggers are outlined.

\section{Color fluctuation effects in proton-nucleus collisions}\label{sec:sec2}
\subsection{Gribov inelastic shadowing}
It was demonstrated by Gribov \cite{Gribov:1968jf} that the nuclear shadowing contribution to the total 
cross section of the hadron-deuteron scattering can be expressed through the diffraction cross section
at $t=0$. Operationally this amounts to the replacement in the Glauber formul\ae{ }of the elastic hN cross 
section at $t\sim 0$ by the sum of elastic and diffractive cross section at $t=0$, leading to an enhancement 
of the multinucleon interactions. For heavier nuclei the Gribov formul\ae{ }involve the coupling 
of the projectile to $N>2$ vacuum exchanges which has to be modeled. 

The contribution of the double scattering to the total hadron-nucleon (hN) cross section is enhanced by a
factor $1+\omega_{\sigma}$, where 
\beq 
\omega_{\sigma}=\left.\slfrac{d\sigma(h N \to  X N)\over dt}{d\sigma(h N \to  h N)\over dt}\right|_{t=o}\,.
\eeq
The relation between the double scattering cross section and the total diffraction cross section can be naturally 
understood in the Good and Walker formalism \cite{Good:1960ba}, which provided the effective realization of the 
Feinberg-Pomeranchuk picture \cite{FeinbergPomeranchuk} of the inelastic diffraction. In this formalism one 
introduces eigenstates of the scattering matrix diagonal in $\sigma$; see Ref. \cite{Blaettel:1993ah} for a review. 
Configurations with different $\sigma_i$ scatter  without interference off two target nucleons contributing in the
case of scattering of two nucleons with strength $\propto \sigma_i^2$ to the shadowing of the total cross section. 
This is the same quantity as in the expression for the total cross section of hadron-nucleon diffraction at $t=0$. 
This interpretation of the Gribov result for the shadowing correction to the total cross section was first given 
by Kopeliovich and Lapidus \cite{Kopeliovich:1978qz}.

%%%%%%%%%%%
\subsection{Distribution over  the strength of interaction}
The fluctuations of strength of interaction arise  naturally in QCD where the strength of interaction 
depends on the volume occupied by color.  In particular, the presence of some small configurations leads 
to fluctuations interacting with a small cross section. So we will refer to these fluctuations as 
color fluctuations. 

In order to describe  the effect of  color fluctuations for a variety of processes it is convenient to  introduce 
the notion of distribution over the strength of interaction, $P_h(\sigma_{tot})$ - the probability for an incoming 
hadron to interact with total cross section $\sigma_{tot}$. The distribution $P_h(\sigma_{tot})$ satisfies two 
normalization sum rules:
\beq\label{cond1}
\int d\sigma_{tot}\,P_h(\sigma_{tot})=1\,,\hspace{0.5cm}
\int d\sigma_{tot}\,\sigma_{tot}\,P_h(\sigma_{tot})=\sigma_{tot}^{hN}\,,
\eeq
and the Miettenen-Pumplin relation \cite{Miettinen:1978jb}
\beq\label{cond3}
\int d\sigma_{tot} \left[\sigma^2_{tot}/(\sigma_{tot}^{hN})^2 - 1 \right]
P_h(\sigma_{tot})=\omega_{\sigma}\,,
\eeq
where $\sigma^{hN}_{tot}$ is the free cross section. Experimentally, $\omega_{\sigma}$ first grows with energy then starts 
dropping at energies $\sqrt{s} \ge$ 100 GeV. There are no direct measurements at the RHIC energy of 200 GeV, but an 
overall analysis indicates that it is of the order 0.25. The first LHC data seem to indicate that inelastic diffraction 
still constitutes a large fraction of the cross section - it is comparable to the elastic cross section, suggesting 
$\omega_{\sigma} \sim 0.2$ at those energies. It is difficult at the moment to ascribe error bars to these numbers. 
However, it is expected that the values of $\omega_{\sigma}$ corresponding to the LHC energies will be soon measured  
with a good precision.

It is worth emphasizing here that these seemingly small values of $\omega_{\sigma}$ correspond to very large fluctuations 
of the interaction strength. For example, if %can 
we consider a simple two component model (equivalent to the 
quasi-eikonal approximation), in which two components are present %exist 
in the projectile wave function with equal probability
and interact with strengths $\sigma^{(1)}_{tot}$ and $\sigma^{(2)}_{tot}$:
\beq
\sigma^{(1)}_{tot}\,=\,\sigma^{hN}_{tot}(1- \sqrt {\omega})\,,\hspace{0.5cm}
\sigma^{(2)}_{tot}\,=\,\sigma^{hN}_{tot}(1+ \sqrt {\omega})\,.
\eeq
Thus for $\omega_{\sigma}=0.25$, we have $\sigma^{(1)}_{tot}/\sigma^{hN}_{tot}=0.5$, 
$\sigma^{(2)}_{tot}/\sigma^{hN}_{tot}=1.5$ and hence $\sigma^{(1)}_{tot}/\sigma^{(2)}_{tot}=3$.

\section{Gribov - Glauber model predictions for fluctuations in the optical approximation}\label{sec:sec3}
In order to illustrate the effects of the color fluctuations and their interplay with the fluctuations of the 
local nuclear density we first consider the optical approximation of the Glauber model where the radius of the 
NN interaction is neglected as compared to the distance between the nucleons. 

Within this model the total inelastic hadron-nucleus cross section $\sigma^{hA}_{in}$ can be written as follows:
\begin{equation}\label{eq7.2}
  \sigma^{hA}_{in} = \int d\Vec{b}  \; \left(1 - \left[1 - x(b)\right]^{A}\right) =
  \sum^{A}_{N = 1} \frac{(-1)^{N + 1} A!}{(A - N)! \,N!}
  \int  d\Vec{b}\, x(b)^{N}\,,
\end{equation}
where $x(b)=\sigma^{hN}_{in}\,T(b)/A$ and normalization $\int d\Vec{b}\,T(b)=A$.

Note that in Eq. (\ref{eq7.2}) nucleon-nucleon correlations in the nuclear wave function are neglected 
as well as the finite radius of the hadron-nucleon interaction; an implementation of correlations in 
the optical limit in the Gribov - Glauber formalism can be found in Refs. \cite{Alvioli:2008rw,Alvioli:2009iw}
and in Ref. \cite{Alvioli:2009ab} within the MC approach and will not be discussed here. Eq. (\ref{eq7.2}) can 
be rewritten as a sum of positive cross sections \cite{Bertocchi:1976bq} as follows:
\begin{equation} \label{eq7.2a} 
  \sigma^{hA}_{in} = \sum^{A}_{N = 1} \sigma_{N},
  \quad \sigma_{N} = \frac{A!}{(A - N)!\, N!} \int
  d\Vec{b}\, x(b)^N \left[1-x(b)\right]^{A-N},
\end{equation} 
where $\sigma_{N}$ denotes the cross section of the physical process in which $N$ nucleons have been involved 
in inelastic interactions with the projectile. Using Eq. (\ref{eq7.2a}), the average number of interactions 
$\left<N\right>$ can be expressed as
\beq \label{eq7.3}
\left<N\right>\,=\,\sum^{A}_{N = 1} N \,\sigma_{N} \biggl/
\sum^{A}_{N = 1} \sigma_{N}=\frac{\sigma^{hN}_{in}}
     {\sigma^{hA}_{in}} \int d\Vec{b}
     \,T(\mathbf{b}) = \frac{A \sigma^{hN}_{in}}
        {\sigma^{hA}_{in}},
\eeq
which coincides with the naive estimate of shadowing as being equal to the number of nucleons shadowed 
in a typical hA inelastic  collision. 

We can include color fluctuations by allowing the inelastic cross section $\sigma_{in}$ to be 
distributed according to a proper distribution, $P_H(\sigma_{in})$:
\begin{equation} \label{fl1}
  \sigma^{hA}_{in} =\int d\sigma_{in} P_H(\sigma_{in}) \int d\Vec{b} 
  \, \left(1 - \left[1 - x(b)\right]^{A}\right)\,, 
\end{equation}
where now $x(b)=\sigma_{in}\,T(b)/A$, and
\begin{equation}\label{fl2}
  \quad \sigma_{N} = \int d\sigma_{in} P_H(\sigma_{in})\frac{A!}{(A - N)!\, N!} \int
  d\Vec{b}\, x(b)^{N} \left[1 - x(b)\right]^{A - N}\,.
\end{equation} 
The probability of collisions with exactly $N$ inelastic interactions in both Glauber model 
and the color fluctuation approximation is simply $R_N= \sigma_N/\sigma^{hA}_{in}$.

Using the equations above we can for example calculate the average number of collisions which is 
given by the same equation as for the Glauber model (Eq. (\ref{eq7.3})), leading to a very small 
(a few \%) change of average $N$, as shown in Table \ref{tabONE}, since the inelastic corrections 
to $\sigma_{in}^{hA}$  are small for a realistic $P_H(\sigma_{in})$; see Ref. \cite{Alvioli:2009iw}
and references therein. The physical reason why the corrections are small is that, in a broad 
range of $b$, the interaction is close to the black limit for all essential values of $\sigma_{in}$, 
so only a small range of (large) $b$ contributes to inelastic shadowing corrections.
At the same time the color fluctuation effect is large for  the variance of the distribution over the number of collisions. 
Eq.(\ref{fl2}) leads to 
\beq 
\left<N(N-1)\right>= A(A-1)\,\frac{\left<\sigma_{in}^2 \right>}{\sigma^{hA}_{in}} \int d\Vec{b}\, T^2(b),
\label{avN2}
\eeq
and hence the variance is equal to
\beq 
\label{varoptical}
\omega_N\equiv  { \left<N^2\right>\over \left<N\right>^2} -1=
\frac{A(A-1)}{\left<N\right>^2}\,\frac{\left<\sigma_{in}^2 \right>}{\sigma^{hA}_{in}}
\int d\Vec{b}\, T^2(b) +{1\over \left<N\right>} -1\,.
\eeq
One can see from Eq.(\ref{varoptical}) that  the variance receives contributions both from the fluctuations 
of the impact parameter and from the fluctuations of $\sigma_{in}$. Using Eqs.(\ref{eq7.3}),(\ref{avN2})
we obtain for the variance in Eq.(\ref{varoptical}) the  value of about 0.46 (RHIC) and 0.51 (LHC).
Numerical values of the different terms in Eq.(\ref{varoptical}) are: 1.26 +0.20 -1 = 0.46 (RHIC) and 
1.38 +0.13 -1 = 0.51 (LHC). 
The account of the color fluctuations practically does not change $\left<N\right>$. It  mainly changes the nominator 
of the first term by the factor $1+\omega_{\sigma}$.\footnote{We assume here that fluctuations for the inelastic and 
total cross sections are similar, cf. discussion before Eq.  (\ref{pinel}).}.  Though this change is rather small,
the strong cancellation between the first and the third terms of Eq. (\ref{varoptical}) strongly enhances the effect 
of color fluctuations. 
  
A more realistic treatment of the color fluctuations taking into account the profile function of the NN 
interactions and small effect of short-range correlations is possible in the MC model described in the 
next section. First, one calculates  the probability $P_N(b)$ shown in Fig. 
\ref{figONE} of having exactly $N$ inelastic interactions at a  given 
impact parameter $b$. Next one can calculate  the 
quantity in Eq.(\ref{varoptical})  by  integrating $P_N(b)$  over the impact parameter: $P_N = 2\pi \int\,b\,db\,P_N(b)$. The results 
are given in Table \ref{tabONE}.
\begin{table}[!ht]%=================================================== Table ONE
  \begin{center}
    \begin{tabular}{|c||c|c|c||c|c|c||} \hline
       & \multicolumn{3}{c||}{Monte Carlo} & \multicolumn{3}{c||}{Optical Model} \\\hline 
      energy/model & \hspace{0.2cm}$\langle N\rangle$\hspace{0.2cm} 
                   & \hspace{0.2cm}$\langle N^2\rangle$\hspace{0.2cm} 
                   & \hspace{0.2cm}$\omega_N$\hspace{0.2cm}
                   & \hspace{0.2cm}$\langle N\rangle$\hspace{0.2cm} 
                   & \hspace{0.2cm}$\langle N^2\rangle$\hspace{0.2cm} 
                   & \hspace{0.2cm}$\omega_N$\hspace{0.2cm}\\\hline
      RHIC, Glauber         &  4.6 & 31.6 & 0.51 & 5.0 &  35.9 & 0.46\\
      RHIC, GG2             &  4.7 & 38.9 & 0.74 & 5.1 &  45.3 & 0.71\\     
      RHIC, GG $P_h(\sigma_{tot})$  &  4.8 & 39.2 & 0.72 & 5.2 &  45.6 & 0.70\\\hline
      LHC, Glauber          &  6.7 & 72.4 & 0.59 & 7.6 &  88.0 & 0.51\\
      LHC, GG2              &  6.8 & 84.2 & 0.80 & 7.8 & 106.2 & 0.75\\
      LHC, GG $P_h(\sigma_{tot})$   &  6.8 & 82.1 & 0.77 & 7.8 & 106.4 & 0.74\\\hline
    \end{tabular}
    \caption{The fluctuations, as defined in Eq. (\ref{varoptical}), calculated both within the MC
      approach and optical model. We used no color fluctuation (Glauber), color fluctuations implemented 
      with the two states model described in the text (GG2) and with the full color fluctuation model 
      (GG $P_h(\sigma_{tot})$) described by the distribution $P_h(\sigma_{tot})$ of Eq. (\ref{psigma}).}
  \label{tabONE}
  \end{center}
\end{table}%========================================================== End Table ONE

A comparison of some of the predictions of the optical approximation of the Glauber model 
and the MC calculations, which take into account finite radius of the NN interaction neglected 
in the optical model, will be given below.

\section{Monte Carlo algorithm for modeling effects of fluctuations}\label{sec:sec4}
We have seen from the analysis of the optical model that fluctuations in the number of wounded nucleons 
originate both from color fluctuations and from fluctuations of the number  of nucleons along the path of 
the projectile.

The event-by-event fluctuations of the number of wounded nucleons due to the fluctuations in the number of 
nucleons at a given impact parameter  are present already on the level of the Glauber model \cite{Alvioli:2009ab}. 
In the case when no fluctuations of $\sigma$ are present, 
$\left< N(\sigma^{hN}_{in})\right>$ is given by Eq. (\ref{eq7.3}). In this case we can write
\beq
\left< N(\sigma^{hN}_{in})^2\right>=\left< N\right>^2 (1+ \omega_{\rho}(\sigma^{hN}_{in}))\,,
\eeq
where $\omega_{\rho}(\sigma^{hN}_{in})$ is the dispersion in the case of no color fluctuations. We found that 
$\omega_{\rho}(\sigma^{hN}_{in})$ drops as a function of
 $\sigma^{hN}_{in}$, as a consequence of the increasing 
number of nucleons in the interaction volume. In the calculations we use the event generator \cite{Alvioli:2009ab}.
This event generator includes short-range correlations between nucleons, however this effect 
leads to a very small correction for the discussed quantity. 
The code also includes a realistic dependence of the probability of the $NN$ interaction on  the relative impact parameter of the projectile 
$\Vec{b}$, 
and the target nucleon $\Vec{b}_j$: $\Vec{b} - \Vec{b}_j$.
The probability of the interaction is expressed 
through the impact factor of the $NN$ elastic amplitude  
\beq\label{gamma}
\Gamma(\Vec{b}-\Vec{b}_j)\,=\,\frac{\sigma^{hN}_{tot}}{4\pi B}\,e^{-(\Vec{b}-\Vec{b}_j)^2/2B}
\eeq
as follows:
\beq
P(\Vec{b},\Vec{b}_j)\,=\,1\,-\,\left[1\,-\,\Gamma(\Vec{b}-\Vec{b}_j)\right]^2\,.
\eeq
Here we used the exponential fit to the elastic cross section $d\sigma/dt\propto \exp(Bt)$.

In order to perform numerical analyses we follow \cite{Guzey:2005tk}, and take the probability 
distribution for $\sigma_{tot}$ as follows:
\beq
\label{psigma}
P_h(\sigma_{tot})\,=\,\rho\,\frac{\sigma_{tot}}{\sigma_{tot}\,+\,\sigma_0}
\,exp\left\{-\frac{\sigma_{tot}/\sigma_0\,-\,1}{\Omega^2}\right\}\,,
\label{model}
\eeq
where $\rho$ is a normalization constant and we have $\sigma_0=$72.5 mb and $\Omega=$1.01 at LHC energies, while 
$\sigma_0=$32.6 mb and $\Omega=$1.49 at RHIC energies. One can verify that the distribution of Eq. (\ref{psigma})
satisfies the sum rules (\ref{cond1}), (\ref{cond3}), with our values 
$\sigma^{hN}_{tot} = \sigma^{NN}_{tot} = 51.95$ mb for RHIC and $\sigma^{hN}_{tot} = \sigma^{NN}_{tot} = 94.8$ mb for 
LHC energies.

When converting from the distribution over $\sigma_{tot}$, $P_h(\sigma_{tot})$, to the distribution over 
$\sigma_{in}$, $P_H(\sigma_{in})$, we used the geometric scaling observation that the $t$-slope of the elastic 
scattering is proportional to $\sigma_{tot}$. So the ratio $\sigma_{in}/\sigma_{tot} = \lambda$ weakly depends 
on the projectile and energy. Hence we take $\lambda= const$, so that we simply have to use a Jacobian 
$1/\lambda$, with 
\beq 
P_H(\sigma_{in}) = P_h(\sigma_{tot}) / \lambda\,,\hspace{0.5cm}\sigma_{in}=\lambda\,\sigma_{tot}\,.
\label{pinel}
\eeq
Indeed in this case $\int d\sigma_{in} P_H(\sigma_{in})$=1 holds as well. This corresponds to 
$B(\sigma_{tot})=B(\sigma^{hN}_{tot})\,\sigma_{tot}/\sigma^{hN}_{tot}$. 

In our numerical studies we used the fluctuation distribution given by  Eq. (\ref{psigma}), 
$\sigma^{NN}_{tot}$ given above and  $B=14$ GeV$^{-2}$ (RHIC), $B=19.38$ GeV$^{-2}$ (LHC).
This parametrization satisfies the $s$-channel unitarity condition $\Gamma(b) \le 1$. 
In our model this condition holds automatically also for the elastic "color-fluctuation"-nucleon 
amplitude. Our algorithm is a natural extension of that of \cite{Alvioli:2009ab} -- where distribution 
over $N$ was calculated in the Glauber model neglecting
effects of color fluctuations.  

\begin{figure}[!htp]
  \centerline{\hspace{0cm}
    \epsfxsize=6.0cm\epsfysize=4.0cm\epsfbox{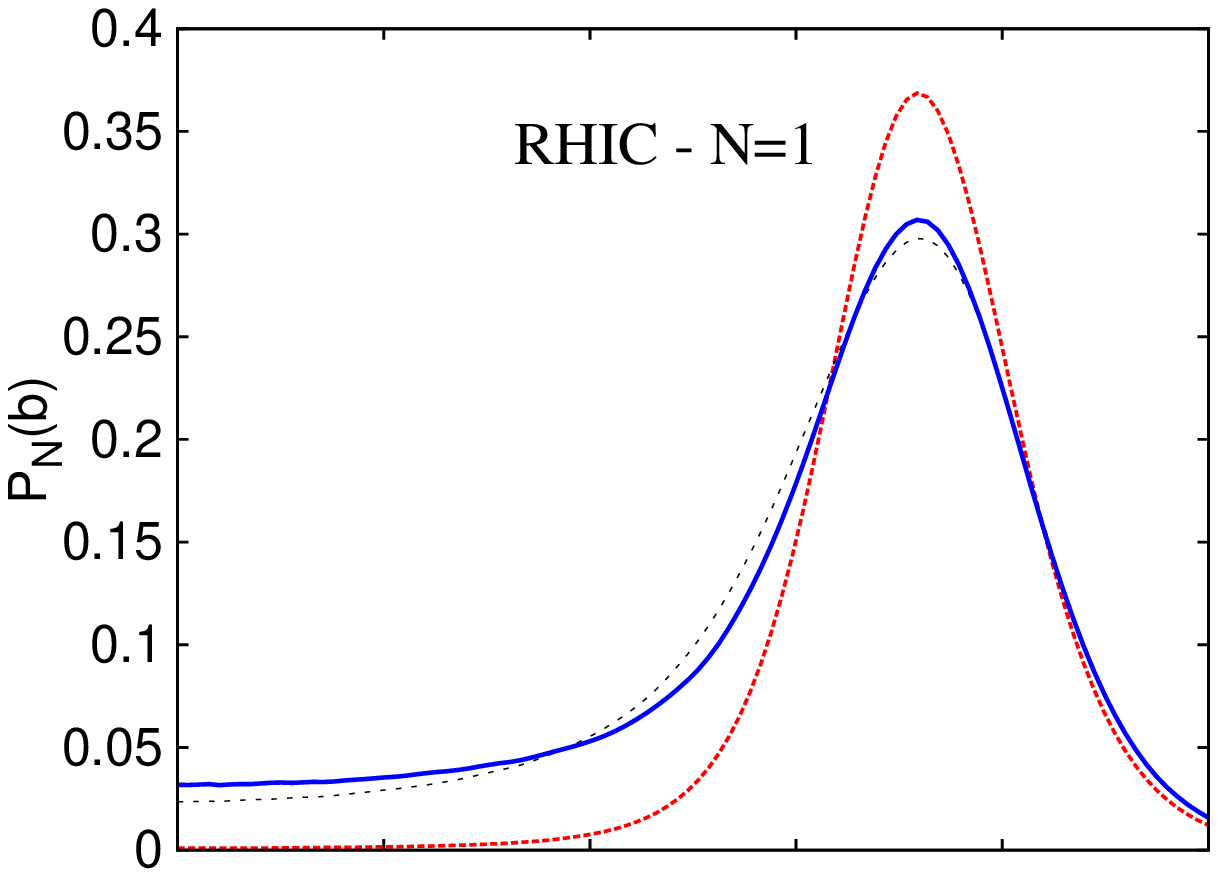}
    \hspace{-0.3cm}
    \epsfxsize=5.7cm\epsfysize=4.0cm\epsfbox{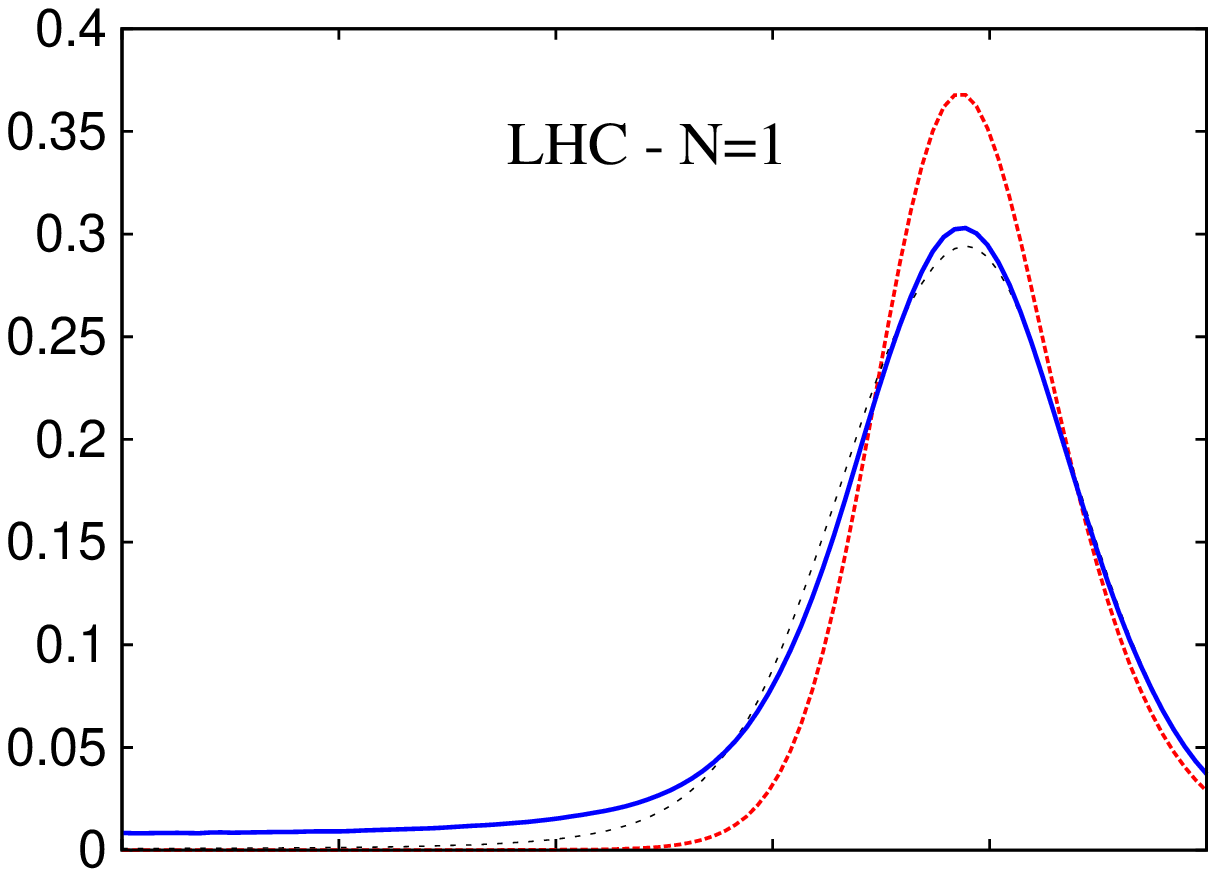}}
  \vskip -0.3cm
  \centerline{\hspace{0cm}
    \epsfxsize=6.0cm\epsfysize=4.0cm\epsfbox{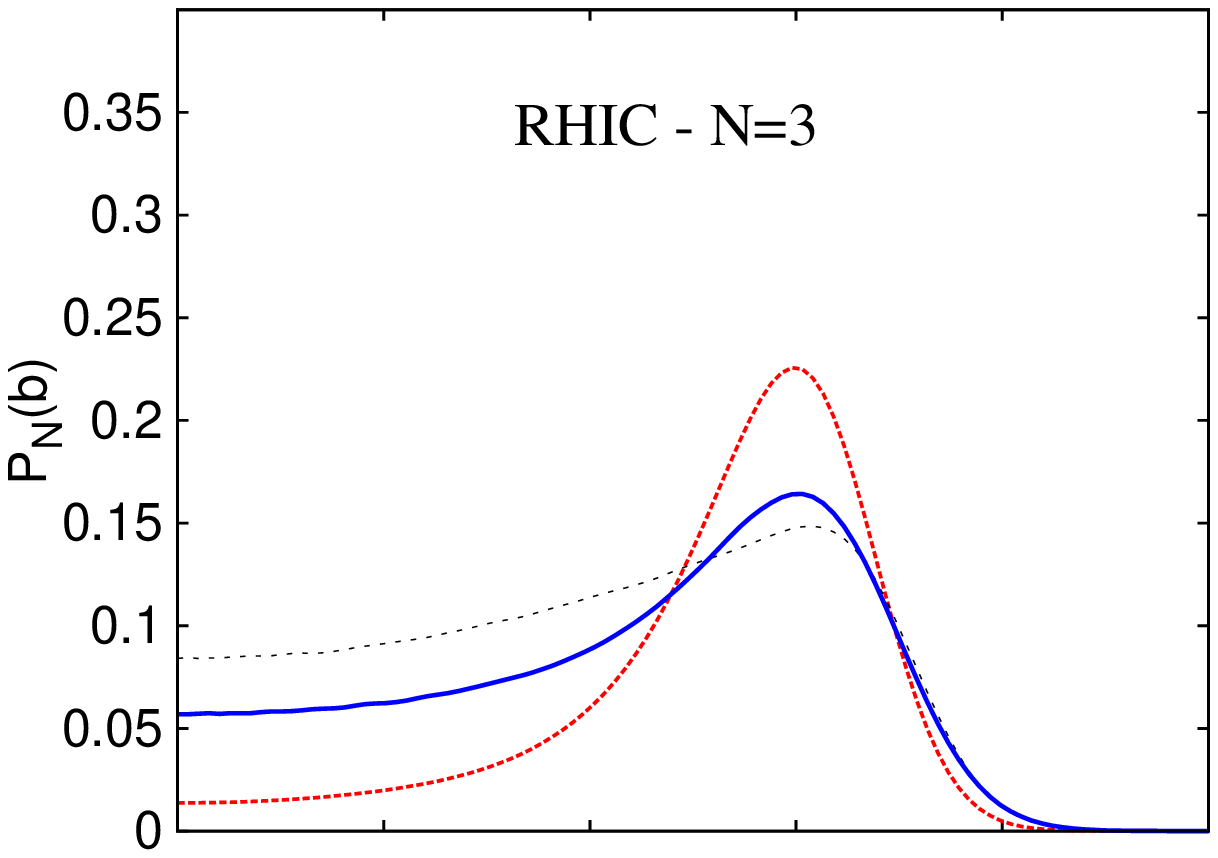}
    \hspace{-0.3cm}
    \epsfxsize=5.7cm\epsfysize=4.0cm\epsfbox{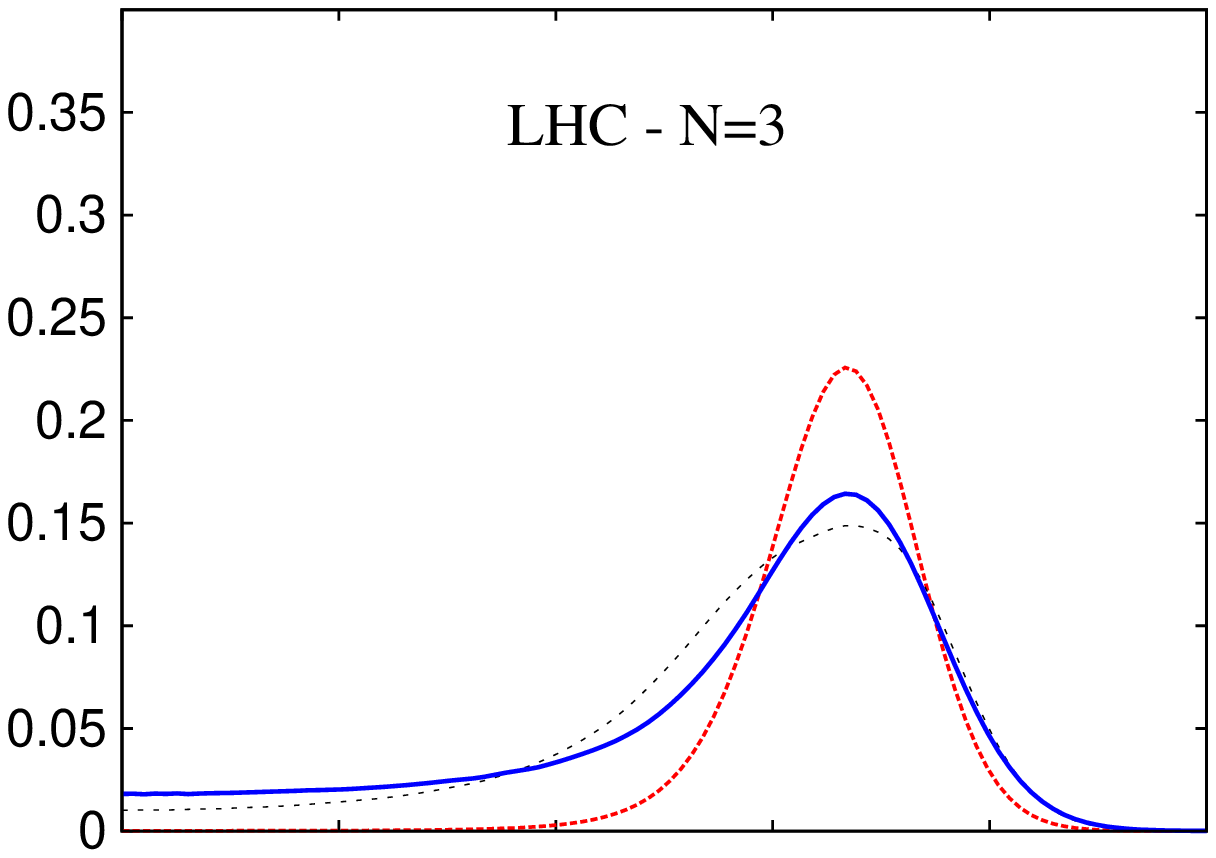}}
  \vskip -0.3cm
  \centerline{\hspace{0cm}
    \epsfxsize=6.0cm\epsfysize=4.0cm\epsfbox{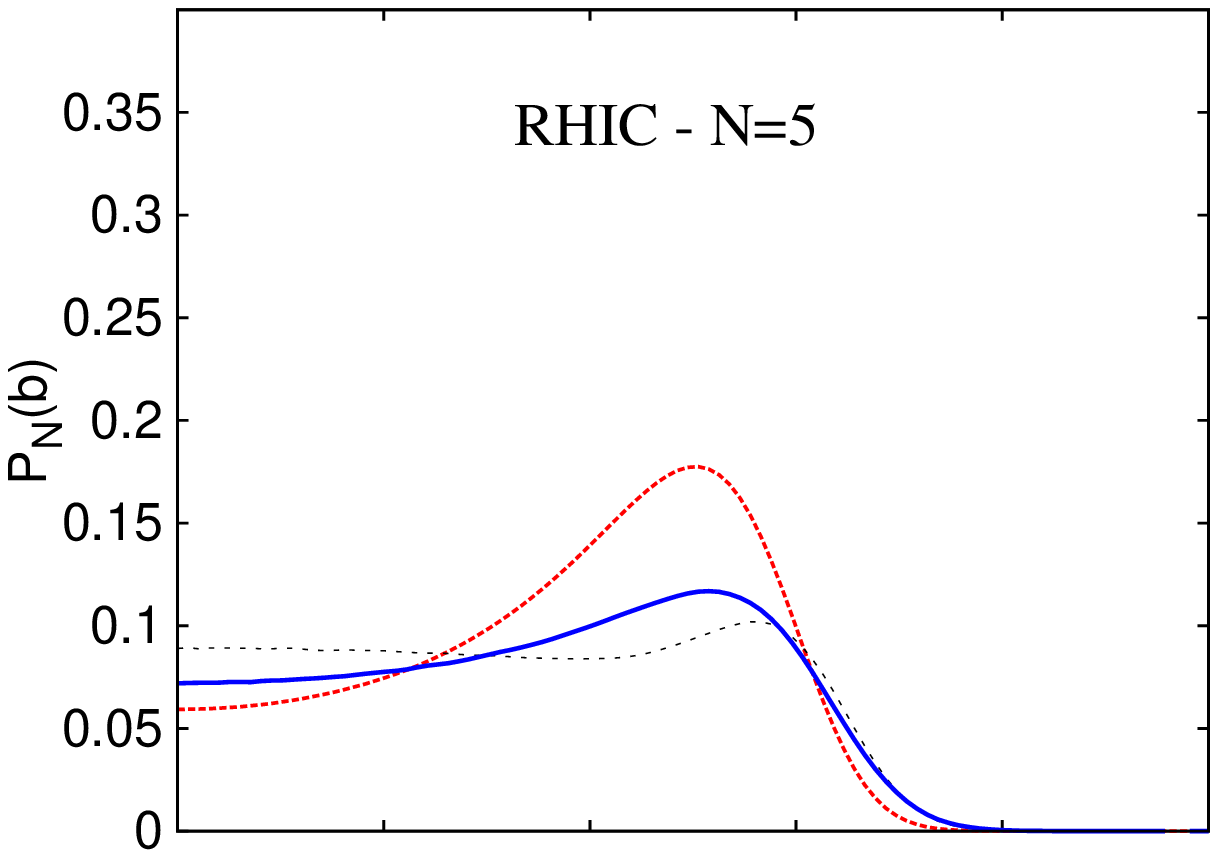}
    \hspace{-0.3cm}
    \epsfxsize=5.7cm\epsfysize=4.0cm\epsfbox{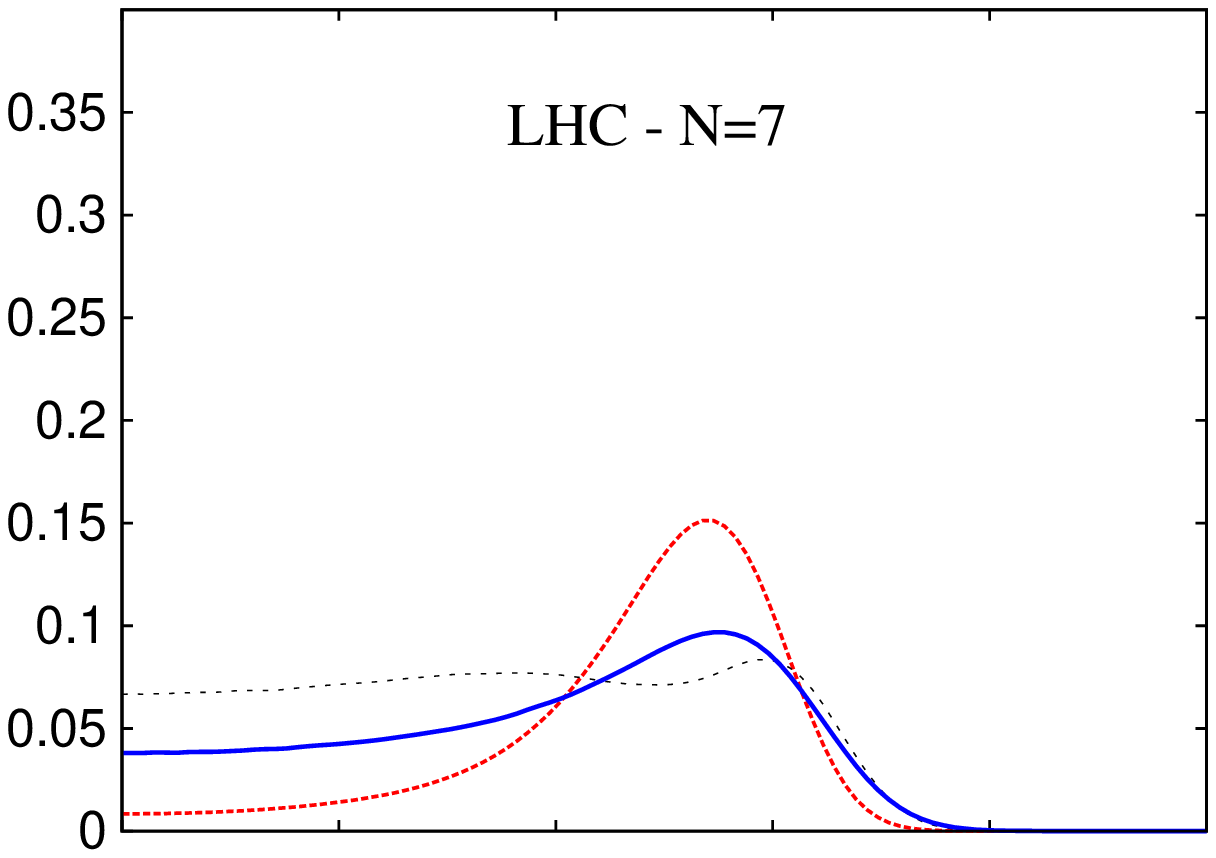}}
  \vskip -0.3cm
  \centerline{\hspace{0cm}
    \epsfxsize=6.0cm\epsfysize=4.0cm\epsfbox{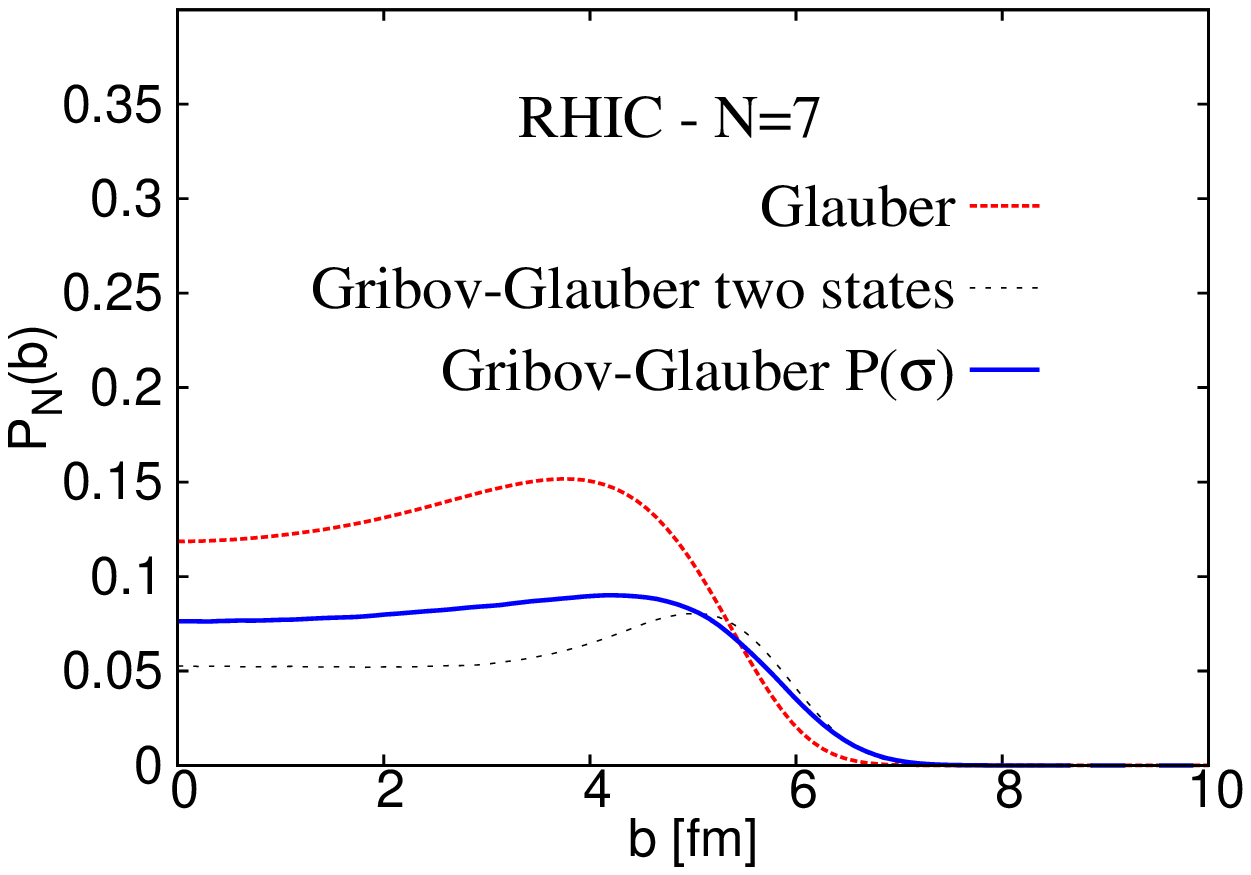}
    \hspace{-0.3cm}
    \epsfxsize=5.7cm\epsfysize=4.0cm\epsfbox{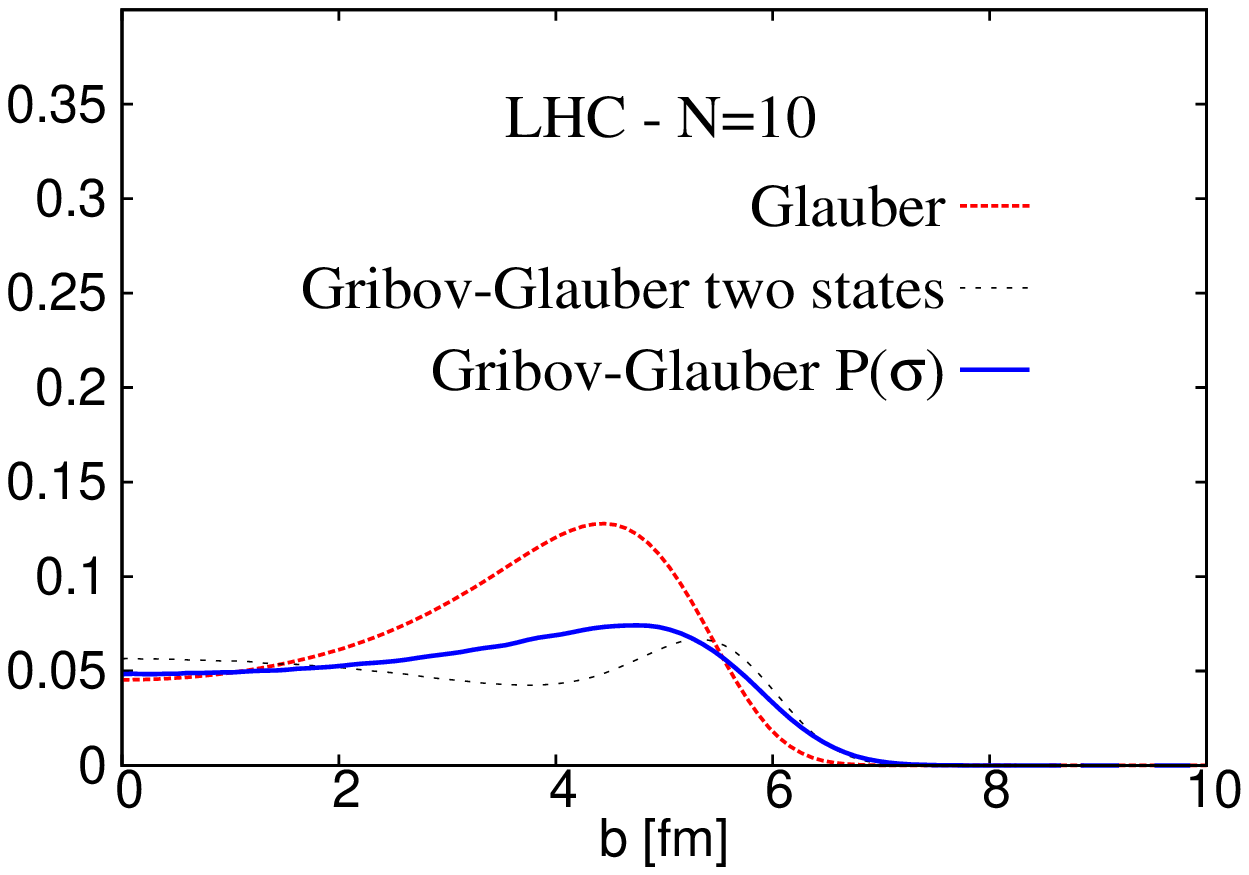}}
  \vskip -0.0cm
  \caption{The probability $P_N(b)$ of having $N$ inelastically interacting (wounded) 
    nucleons in a pA collision, vs. impact parameter $b$, when using simple Glauber 
    (red curves), a two states model (black curves) and a distribution $P_h(\sigma_{tot})$
    (blue curves); \textit{cf.} Eq. (\ref{psigma}). The $P_N(b)$'s are obtained by 
    extension of the MC code of Ref. \cite{Alvioli:2009ab} to include color
    fluctuations.
    Top row shows $P_{N=1}(b)$; the remaining panels correspond to $N = \langle N\rangle$ 
    and $N = \langle N\rangle\pm 0.5\langle N\rangle$.
    $\langle N\rangle$ is taken as 5 and 7 for RHIC and LHC energies, respectively 
    (\textit{cf.} Table \ref{tabONE}).
  }\label{figONE}
\end{figure}

Since the contributions of states with different $\sigma$ do not interfere, the probability $P_N(b)$ to 
have exactly $N$ inelastic interactions at given $b$ is\footnote{In this treatment we neglect 
small
contributions of incoherent diffractive processes $pA \rightarrow XA^{\star}$, which mostly contribute 
to $P_1(b)$.}
\beq
\label{prob04}
P_N(b)\,=\,\int d\sigma_{tot}\,P_h(\sigma_{tot})\,P_N(b;\sigma_{tot}),
\eeq
where $P_N(b;\sigma_{tot})$ is calculated using the procedure of Ref. \cite{Alvioli:2009ab} for fixed
$\sigma^{hN}_{tot}$ in the Glauber model. Including color fluctuations results in a substantially broader 
distribution over $b$ of the probability $P_N(b)$ of having exactly $N$ interactions for a given impact 
parameter $N$, as shown in Fig. \ref{figONE}. The two-component model gives  the distributions pretty 
close to the distributions including full fluctuations. $P_N(b)$ are obviously normalized so that  
$\sum_N \int d\Vec{b} P_N(b) = \sigma^{hA}_{in}$. The calculations of Table \ref{tabONE} 
have been performed integrating the quantities of Fig. \ref{figONE} over the impact parameter: 
$P_N=\int d\Vec{b} P_N(b)$; $\langle N\rangle = \sum_N N P_N / \sum_N P_N$;
$\langle N^2\rangle = \sum_N N^2 P_N / \sum_N P_N$.

\begin{figure}[!htp]
  \centerline{\hspace{0cm}
    \epsfysize=4.3cm\epsfbox{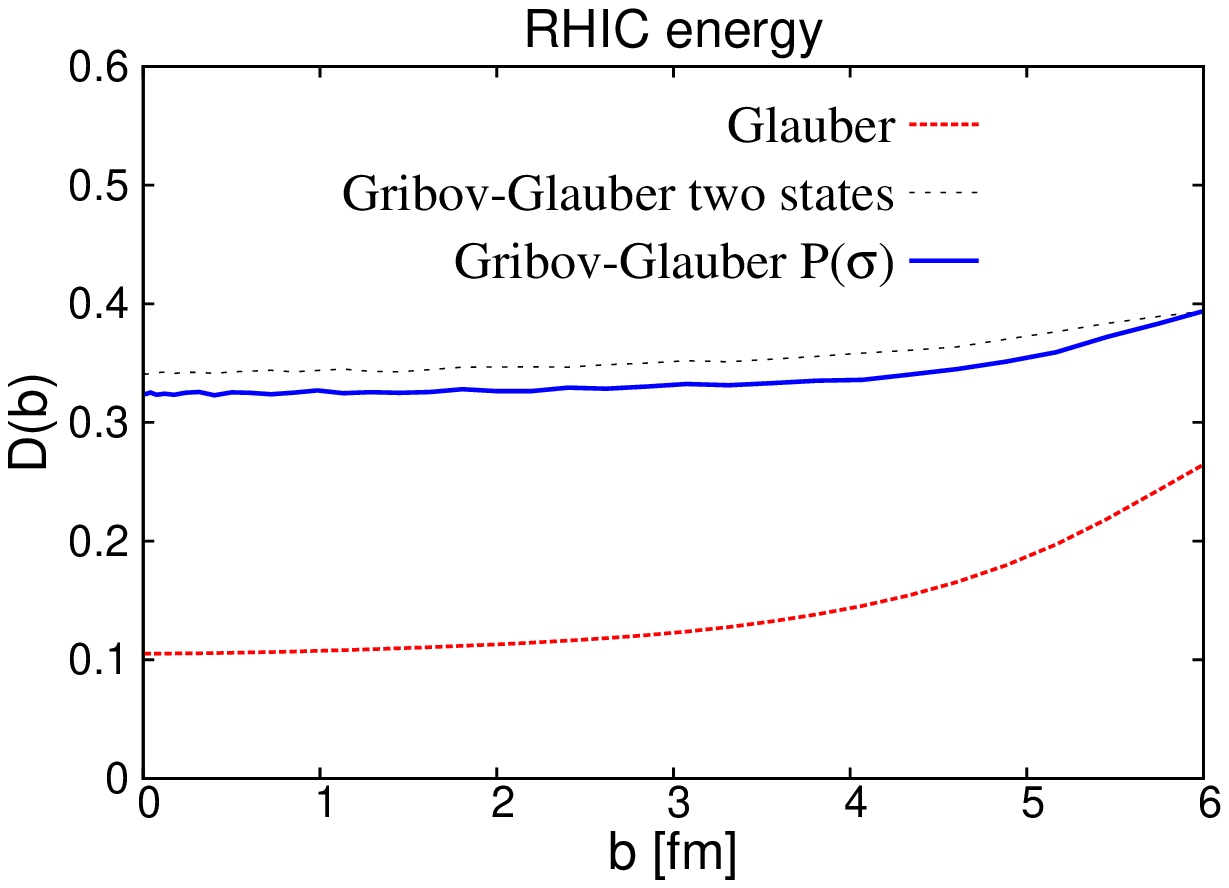}
    \hspace{-0.5cm}
      \epsfysize=4.3cm\epsfbox{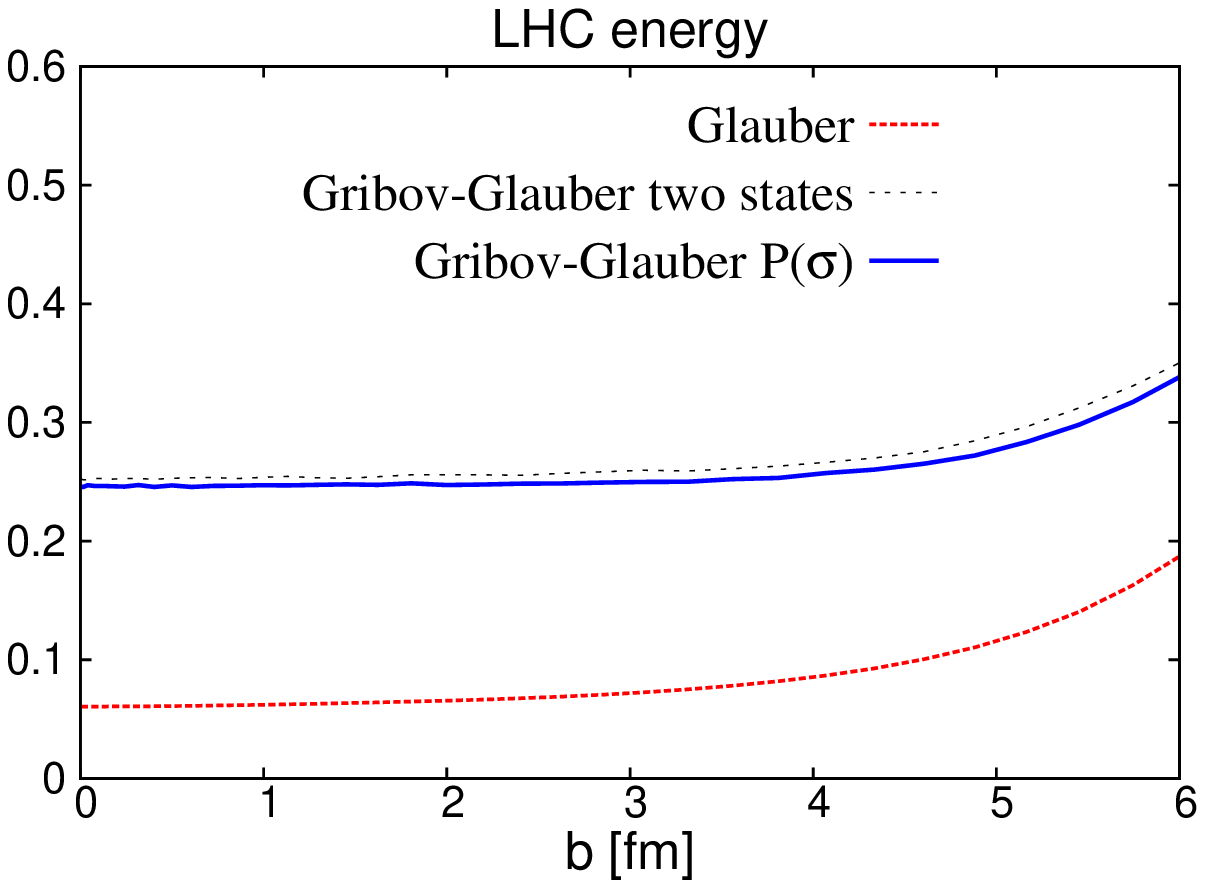}}
  \caption{Effect on fluctuations of the dispersion, Eq. (\ref{dispersion}), when using a 
    distribution of $\sigma_{tot}$ with two values of the cross section with equal probability 
    and with $P_h(\sigma_{tot})$ given by Eq.(\ref{psigma}), for realistic parameters 
    corresponding to RHIC (left) and LHC (right) energies.}
  \label{figTWO}
\end{figure}

Another quantity which characterizes the effects of spatial and color fluctuations is dispersion of the 
number of interactions at a given impact parameter, $b$. To illustrate the expected pattern let us first 
consider   the case of small $b$ and large $A$, when the probability of having at least one inelastic 
interaction is 1. In this case $\langle N\rangle = T(b) \sigma_{in}$, hence the dispersion of the 
distribution over $N$ including both effects can be calculated as follows:
\beq
\label{apprx1}
\left< N^2\right>=\int d\sigma_{in} \,P_H(\sigma_{in})  \left< N\right>^2  
\left({ \sigma_{in}\over \left< \sigma_{in}\right>}\right)^2(1+ \omega_{\rho}(\sigma_{in}))\,.
\eeq
Now we can calculate the total dispersion. The first term in $(1+ \omega_{\rho})$ simply gives $\omega_{\sigma}$. 
The second term takes into account the dependence of $\omega_{\rho}$ on the fluctuating $\sigma_{in}$:
\beq
\label{apprx2}
\omega_{tot}= \omega_{\sigma} + \int  d\sigma_{in} \,P_H(\sigma_{in}) 
\left({ \sigma_{in}\over \left< \sigma_{in}\right>}\right)^2\omega_{\rho}(\sigma_{in})\,.
\eeq
Since the integral in the second term is dominated by $\sigma_{in}  > \sigma^{hN}_{in}$, for which $\omega_\rho$ is 
smaller than in correspondence of the average value of $\sigma_{in}$, $\sigma^{hN}_{in}$, Eq. (\ref{apprx2})
leads to a dispersion somewhat smaller that $\omega_{\sigma}+\omega_{\rho}(\sigma^{hN}_{in})$. This is consistent 
with the pattern we find in the numerical calculation presented in Fig. \ref{figTWO} for \beq\label{dispersion} 
D(b)\,=\,\frac{\langle N^2\rangle_b\,-\,\langle N\rangle^2_b}{\langle N\rangle^2_b}\,,
\eeq
$\langle N\rangle_b = \sum_N N P_N(b) / \sum_N P_N(b)$ and 
$\langle N^2\rangle_b = \sum_N N^2 P_N(b) / \sum_N \int P_N(b)$. 
One can see that for RHIC and LHC energies the dominant effect comes from color fluctuations. 
Moreover, the two states approximation gives the result which is very close to the calculation 
with full $P_h(\sigma_{tot})$, so the two states model can be used to simplify modeling of color 
fluctuation effects.

\begin{figure}[!hbp]
  \centerline{\hspace{0cm}
    \epsfysize=4.3cm\epsfbox{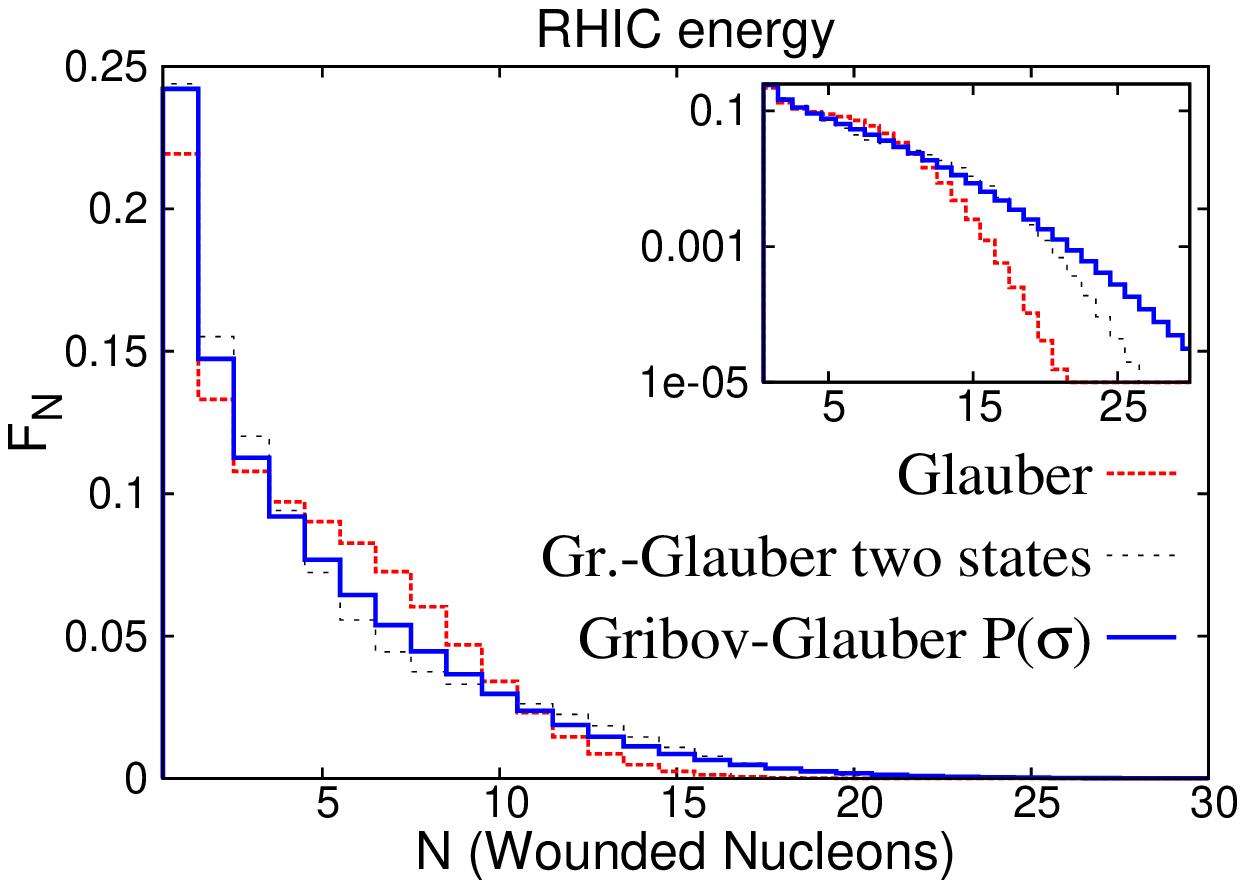}
    \hspace{-0.5cm}
    \epsfysize=4.3cm\epsfbox{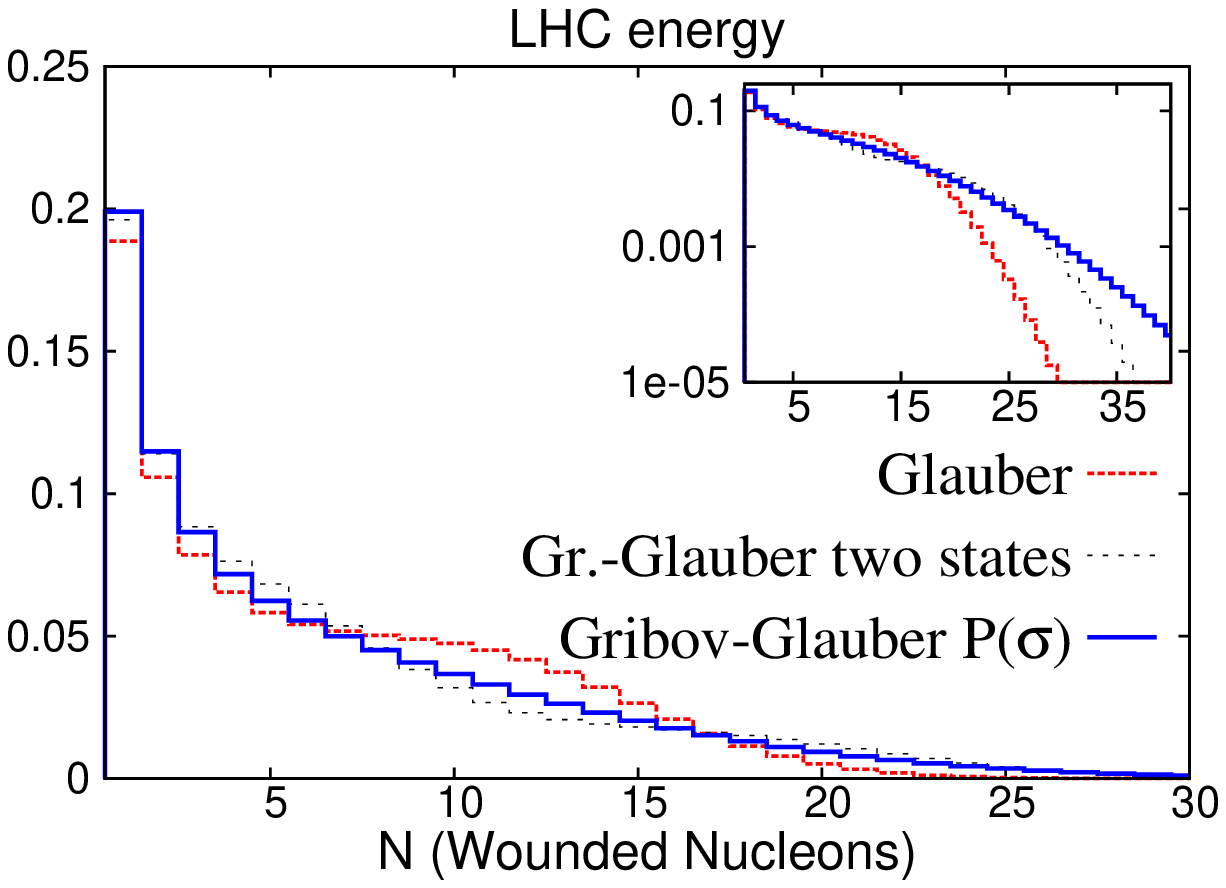}}
  \caption{Effect of the event-by-event fluctuating values of $\sigma_{tot}$, 
    for RHIC (left panel) and LHC energies (right panel) on the number of wounded
    nucleons, calculated as $F_N = \int d\Vec{b}\,P_N(b) / \sigma^{hA}_{in}$. Red 
    curves show the results 
    obtained with the usual Glauber calculation with fixed cross section, black 
    curves correspond to calculations with the two-component $\sigma_{tot}$ model
    and blue curves correspond to calculations with fluctuating cross section
    with $P_h(\sigma_{tot})$ distribution. The insets show the same quantities in
    logarithmic scale.}
 \label{figTHREE}
\end{figure}

The large variance of the distribution leads to a much wider distribution over $N$ than in 
the Glauber model, as shown in Fig. \ref{figTHREE}. The figure shows the quantities 
$F_N = \int d\Vec{b}\,P_N(b) / \sigma^{hA}_{in}$; the same quantities are plotted in logarithmic 
scale in the insets, and one can see that the color fluctuations produce a much stronger large 
$N$ tail. 
Among other things, this implies that selection of events which in the Glauber model correspond 
to very central impact parameters actually gets a significant contribution from pretty large 
impact parameters -- for example, in the two component model discussed above the collisions at 
impact parameter $b$ satisfying the condition $T(b)/T(0) = 1 / (1+ \sqrt {\omega})$ with a 
% ======================================================================== Fig ONE
%% =================================================================== End Fig ONE
% ======================================================================== Fig TWO
%% =================================================================== End Fig TWO
probability of 1/2 generates the same number of wounded nucleons as average number of wounded 
nucleons at $b=0$.  For $\omega=0.25$ we have $1 / (1+ \sqrt {\omega})=0.67$ and this corresponds 
to $b\simeq 4.58$ fm. 
  
\begin{figure}[!hcp]
  \centerline{\hspace{0cm}
    \epsfysize=4.3cm\epsfbox{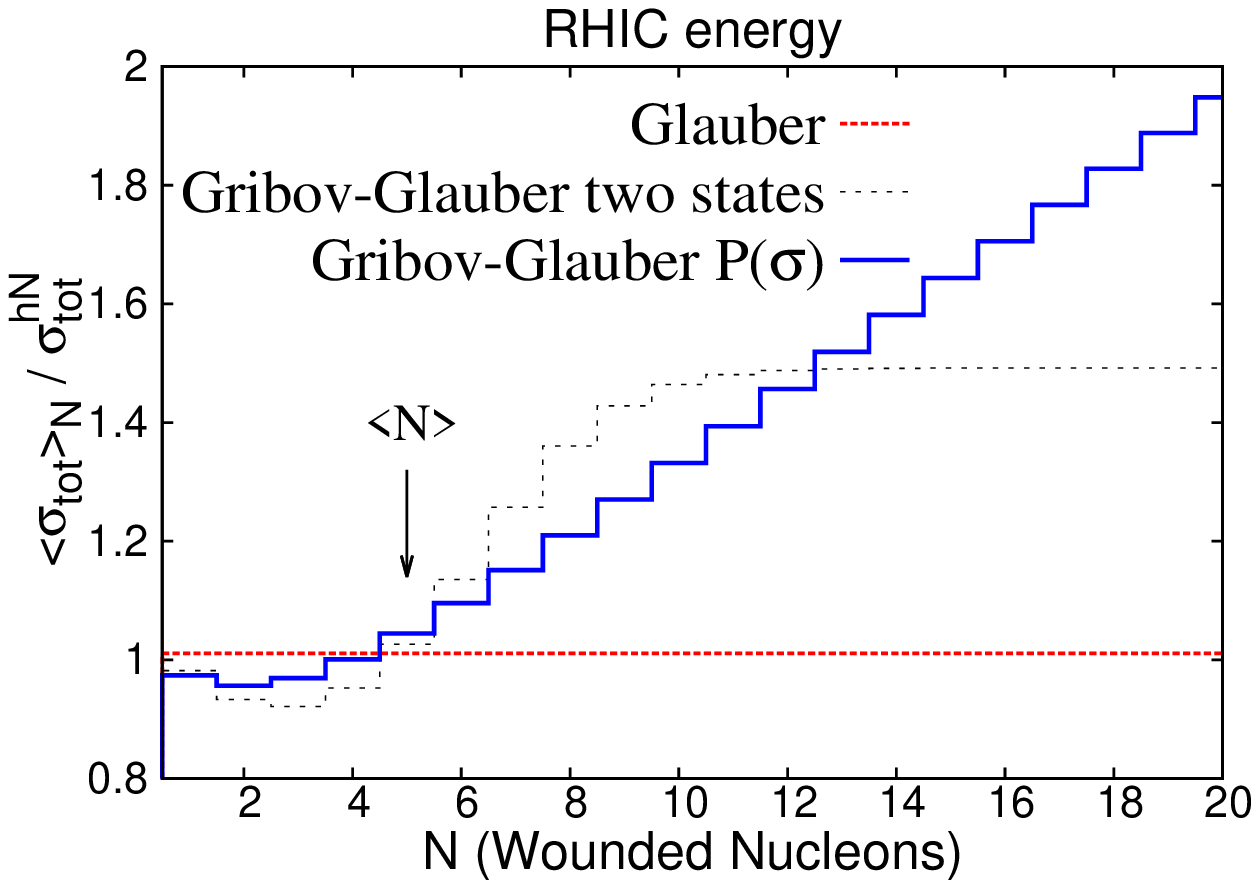}
    \hspace{-0.5cm}
    \epsfysize=4.3cm\epsfbox{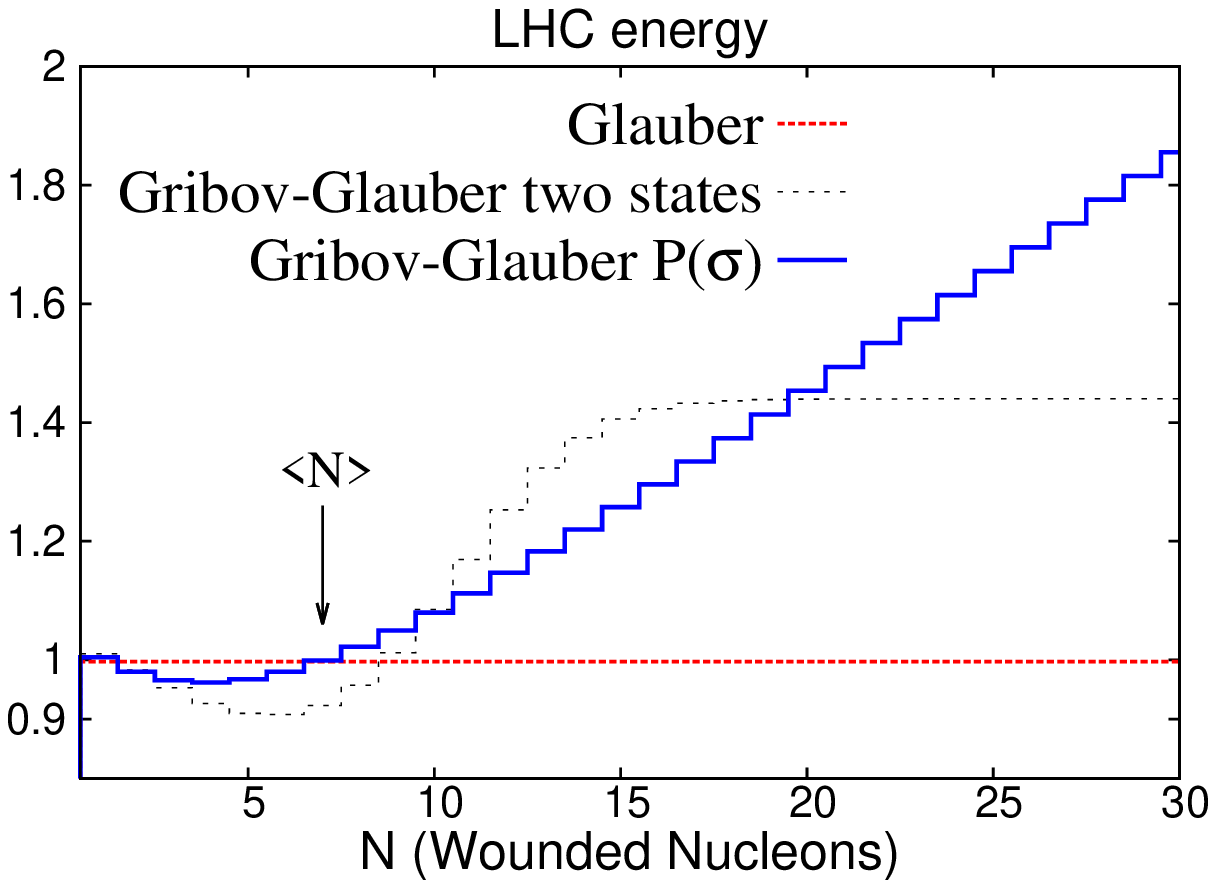}}
  \caption{Effect of fluctuations on the event-by-event fluctuating values of $\sigma_{tot}$, 
    for RHIC and LHC energies.}
  \label{figFOUR}
\end{figure}

An important implication of the broad distributions over $N$ which is mostly due to fluctuations of the strength 
of the interaction is that selection of large $N$ also selects configurations in the projectile nucleon with 
cross section larger than average. To illustrate this trend within our MC, let us consider the average $\sigma_{tot}$ 
for events with a given number $N$ of wounded nucleons. Denoting the probability to have exactly $N$ wounded 
nucleons $P_N=\int d\Vec{b}P_N(b)$ and using Eq.(\ref{prob04}), we can write
% ======================================================================== Fig THREE
%% =================================================================== End Fig THREE
\beq
\label{prob05}
\frac{\langle\sigma_{tot}\rangle_N}{\sigma^{hN}_{tot}}\,=\,\frac{1}{\sigma^{hN}_{tot}}
\,\frac{\int d\sigma_{tot}d\Vec{b}\,\sigma_{tot}\,P_h(\sigma_{tot})\,P_N(b;\sigma_{tot})}
{\int d\sigma_{tot}d\Vec{b}\,P_h(\sigma_{tot})\,P_N(b;\sigma_{tot})}\,.
\eeq
The results of the calculation are presented in 
 Fig. \ref{figFOUR}. One can see  that selecting  $N \gg \langle N\rangle$ leads to a significant enhancement 
of the contribution of configurations which have interaction strength larger than average. For small $N$ average 
$\langle \sigma_{tot}\rangle_N$ is below $\sigma^{hN}_{tot}$, but the effect is relatively small especially for $N=1$ 
where very peripheral collisions contribute which are not sensitive to the fluctuations. A natural source of large 
$\sigma$'s are configurations of  larger than average transverse size.   One can expect that the gluon field 
is enhanced in these configurations while the distribution in $x$ -- the light-cone fraction carried by partons of 
the projectile -- is softer for large $x$ leading to a correlation between the distribution over $N$ and distribution 
over $x$ of a hard collision. 
% ======================================================================== Fig FOUR
%% =================================================================== End Fig FOUR

Matching the number of wounded nucleons to the physical observables is certainly a challenging problem in view 
of fluctuations of the impact parameter in the collisions. A model independent treatment of this problem would 
require a study of pA collisions for different nuclei. Still the central multiplicity appears to be a good observable 
even in the presence of the color fluctuations.  Indeed in the soft interaction dynamics the hadron multiplicity for 
central rapidities, $ y_{c.m.} \sim 0$, does not depend on $\sigma^{hN}_{tot}$, as it is determined by the density of 
partons in a single Pomeron ladder. Hence the hadron multiplicity for $ y_{c.m.} \sim 0$ should be about the same
for different fluctuations. Also the first studies of the pA collisions at the LHC indicate that to a good 
approximation the hadron multiplicity for $p_t\ge$ 1 GeV is proportional to  the number of wounded nucleons  
calculated in the Glauber model \cite{ALICE:2012mj}. Hence we expect that selecting events with the 
$y_{c.m.} \sim 0$ hadron multiplicities: $M / \langle M\rangle \ge 2.5 $ should  select configurations in 
the projectile significantly larger than average ones (cf. Fig. \ref{figFOUR}b) with significantly different
parton distributions.

Correspondingly, a trigger for configurations of smaller than average size would lead to a more narrow distribution 
in $N$. One such possibility is to select as a trigger a hard process in which a parton of the proton with
$x_p > 0.6$ is involved. One may expect that in 
this case one selects quark-gluon configurations without $q\bar q$ pairs and significantly screened gluon field, 
leading to $\sigma_{in}$ significantly smaller than average and hence  a strong suppression of large $N$ tail 
\cite{Frankfurt:1985cv}. Such measurements appear to  be feasible using the data collected in the 2013 $pA$ run
at the LHC in which a significant number of events with large $x_p$ should have been collected. Since this kinematics 
(for the current LHC detectors) corresponds to very large $p_T$'s of the jets, one expects that for the inclusive 
cross section impulse approximation would work very well.
Hence it would be possible to avoid issues of the final / initial state interactions
and nuclear shadowing in interpreting these data.  

A convenient quantity to  study these effects experimentally would be a measurement of the distribution over $x_p$ 
for different classes of hard collisions at fixed $x_A$ normalized to the distribution in the inclusive $pA$ scattering. 
A large effect is expected for the central collisions where the hard cross section  should be suppressed for large $x_p \geq 0.2 \div 0.3$
and enhanced for  $x \le 0.05$. 

Note that such a measurement  among other things would allow to test in an unambiguous way the explanation of 
the EMC effect at large $x$ as due to the dominance  of the smaller than average size configurations in nucleon 
at $x\ge 0.6$; for a recent review see Ref. \cite{Frankfurt:2012qs}.

\begin{figure}[!htp]
  \centerline{\hspace{0cm}
    \epsfysize=4.3cm\epsfbox{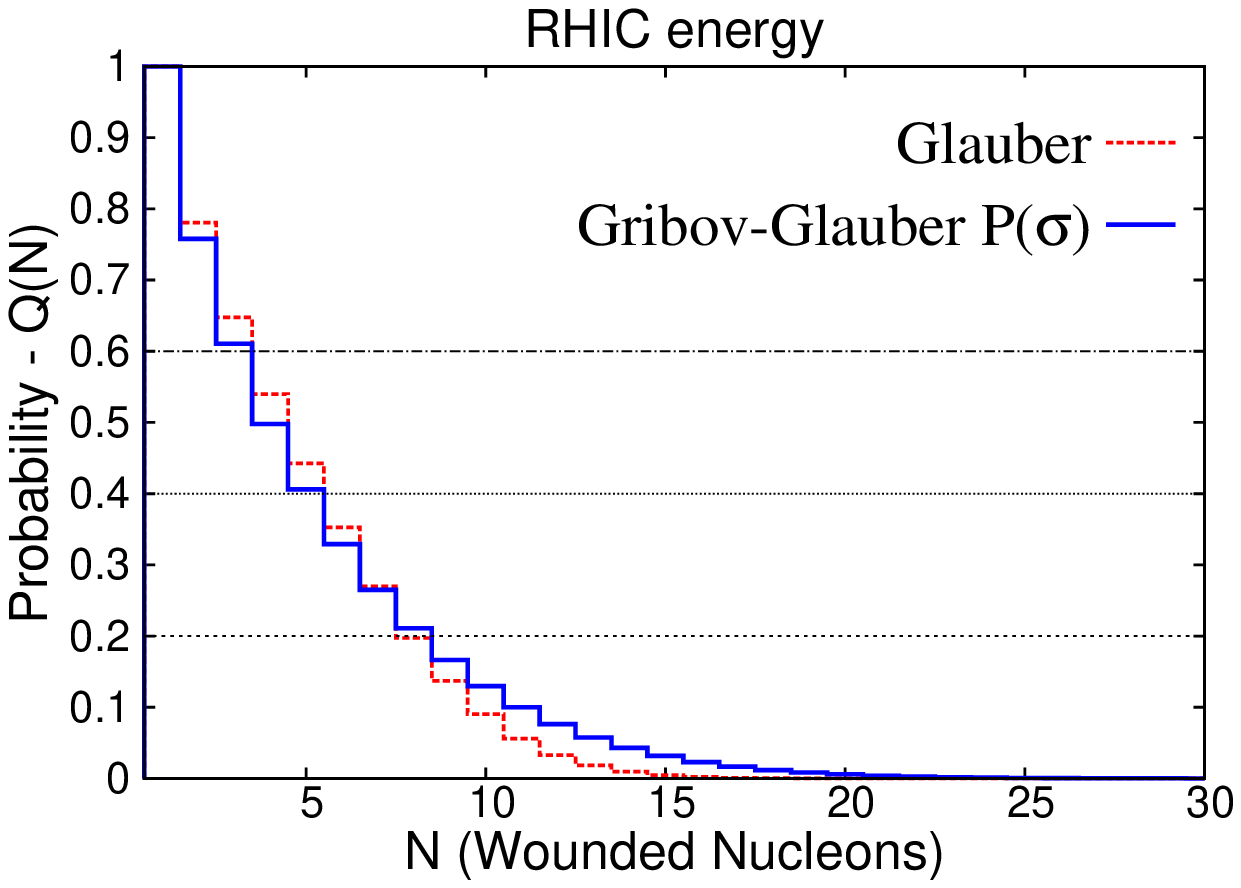}
    \hspace{-0.4cm}
    \epsfysize=4.3cm\epsfbox{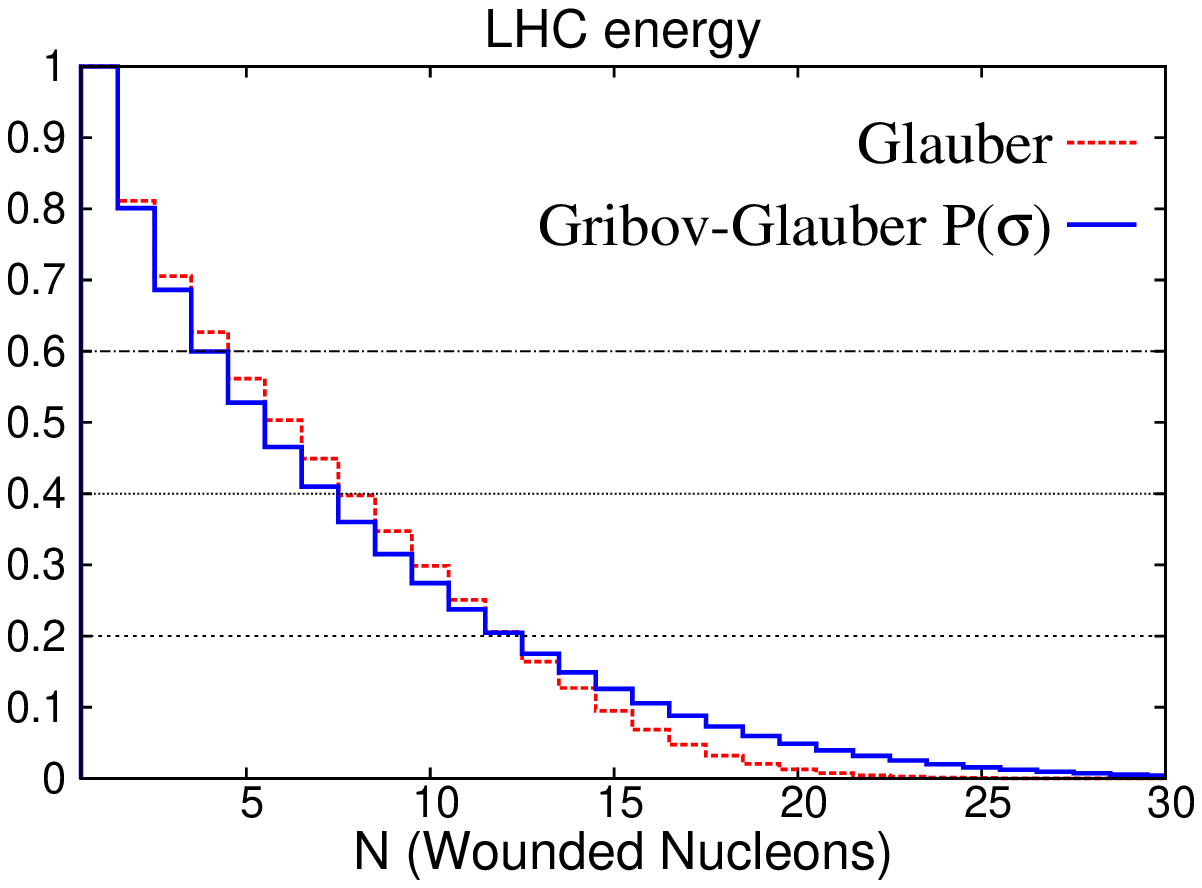}}
  \caption{Fraction of inelastic cross section plotted as a distribution over impact
    parameter as defined in Eq. (\ref{qun}). Horizontal lines at 0.2, 0.4 and 0.6
    correspond to the experimental definition of 20\%, 40\% and 60\% 
    centrality, respectively.}
  \label{figFIVE}
\end{figure}

\begin{figure}[!htp]
  \centerline{\hspace{0cm}
    \includegraphics[width=6.2cm, height=3.8cm]{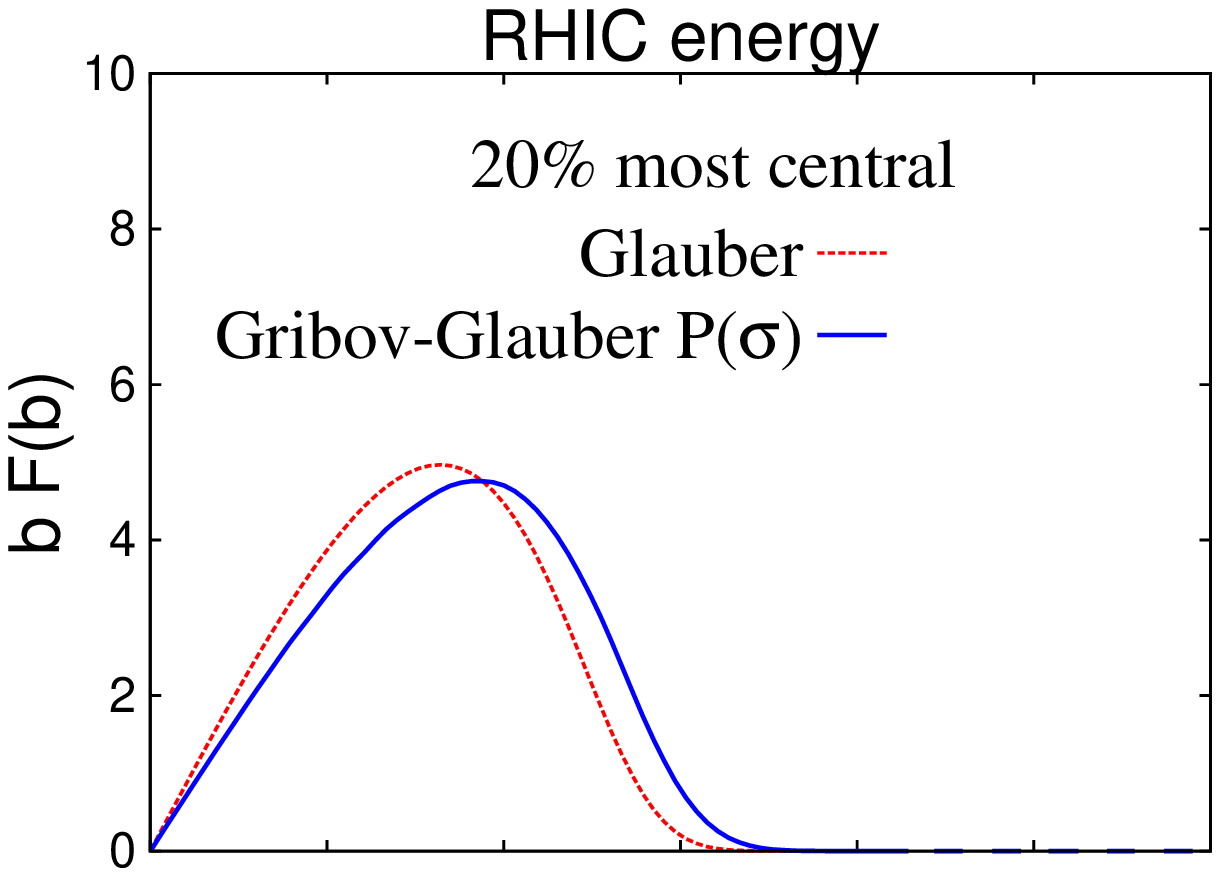}
    \hspace{-0.1cm}
    \includegraphics[width=5.8cm, height=3.8cm]{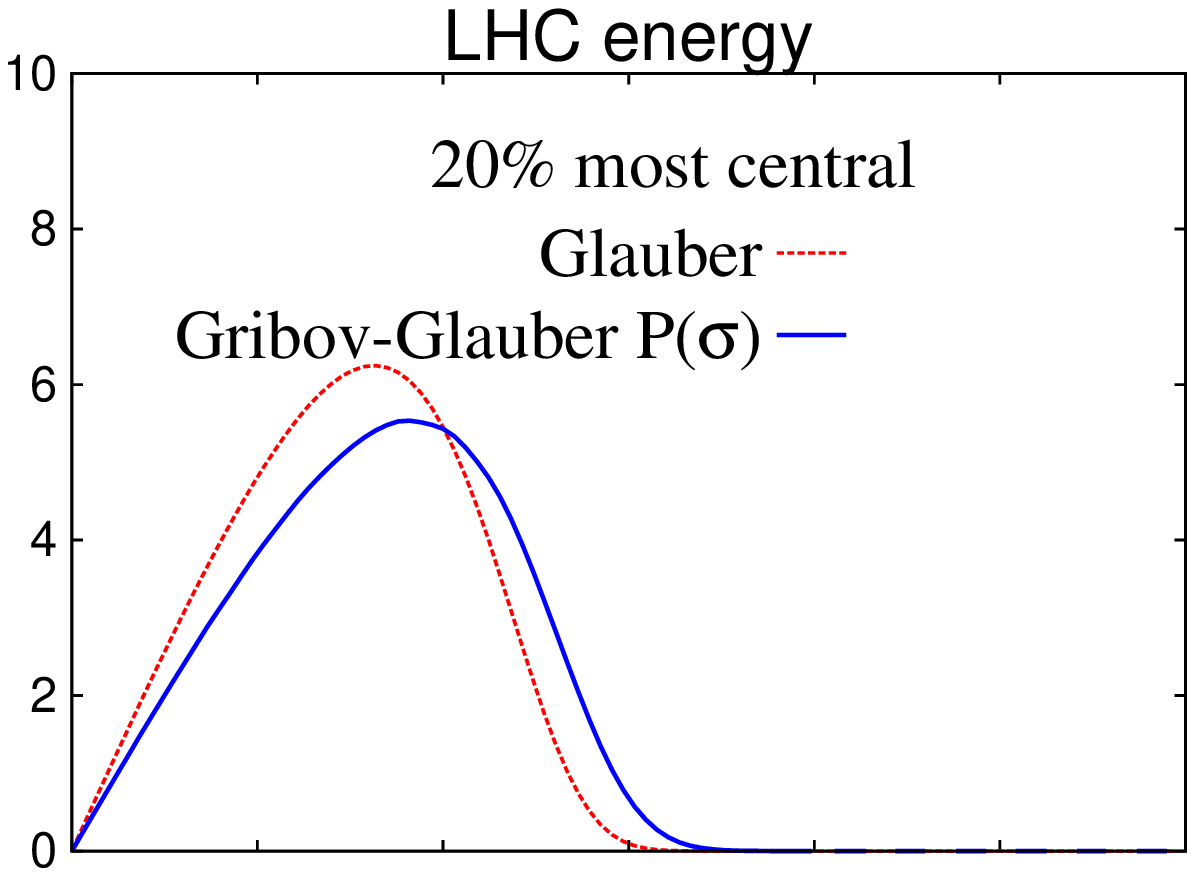}}
  \vskip -0.275cm
  \centerline{\hspace{0cm}
    \includegraphics[width=6.2cm, height=3.8cm]{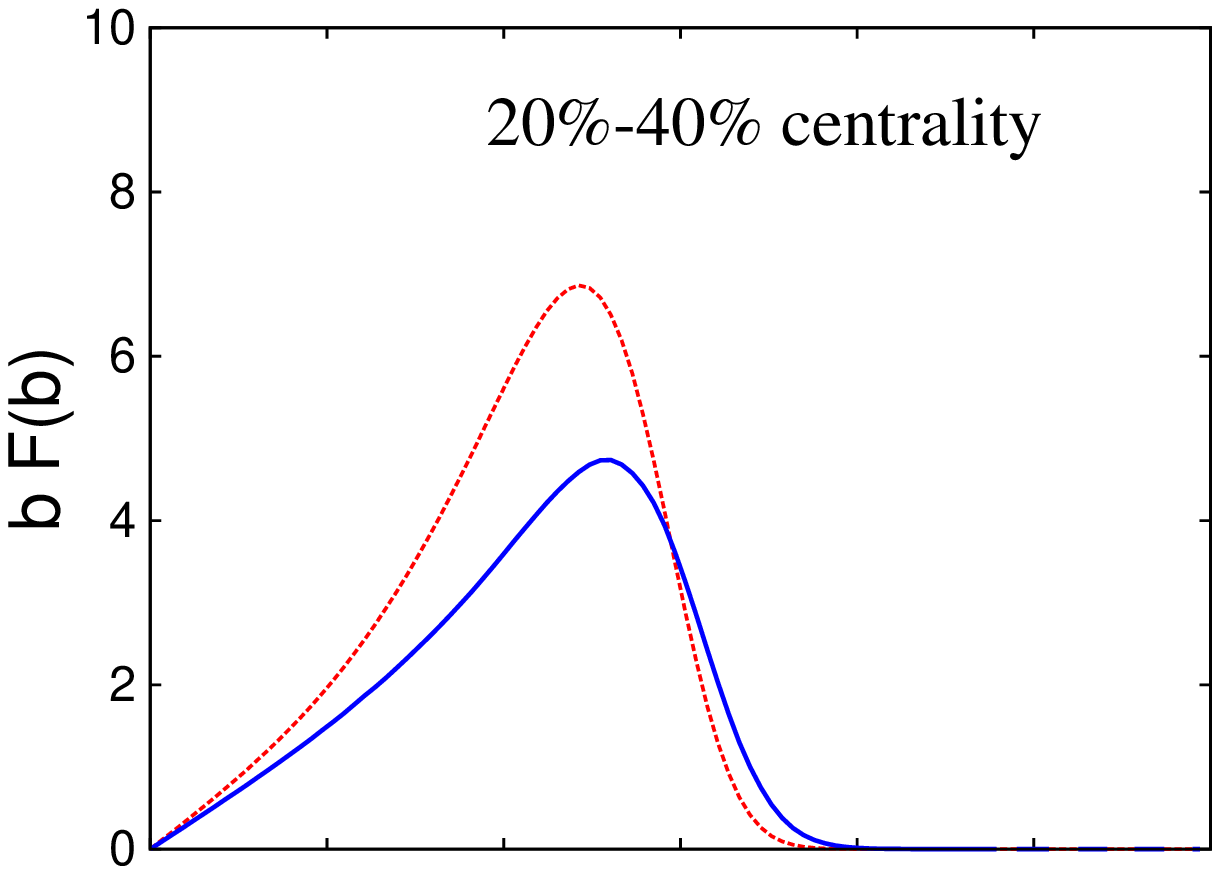}
    \hspace{-0.1cm}
    \includegraphics[width=5.8cm, height=3.8cm]{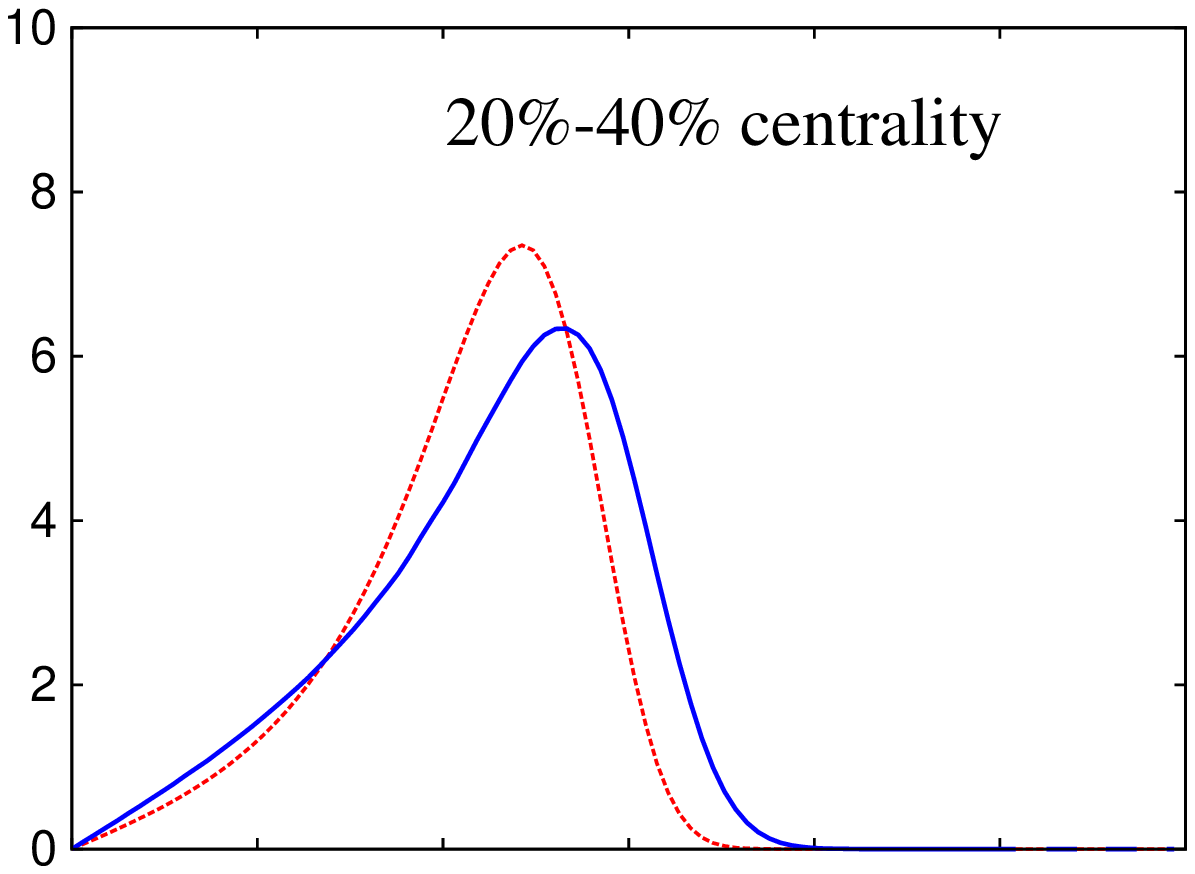}}
  \vskip -0.275cm
  \centerline{\hspace{0cm}
    \includegraphics[width=6.2cm, height=3.8cm]{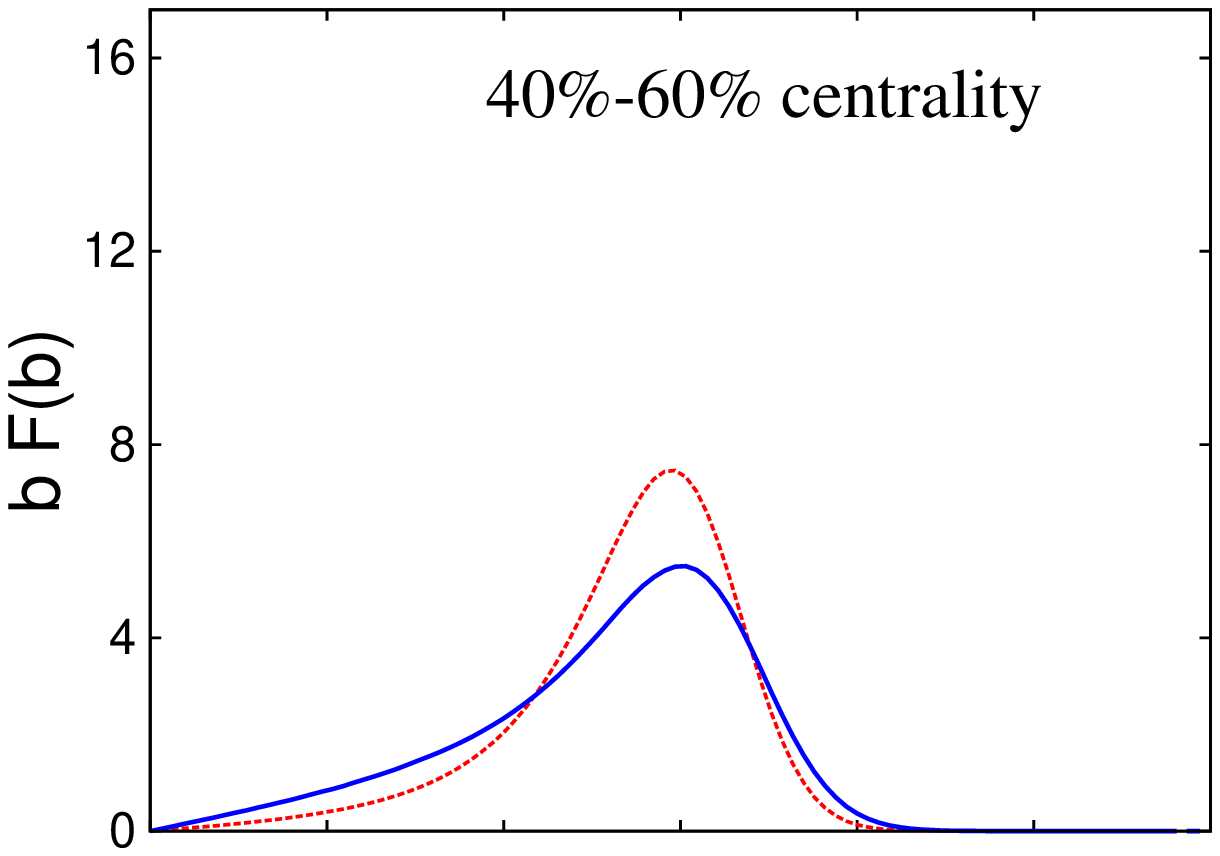}
    \hspace{-0.1cm}
    \includegraphics[width=5.8cm, height=3.8cm]{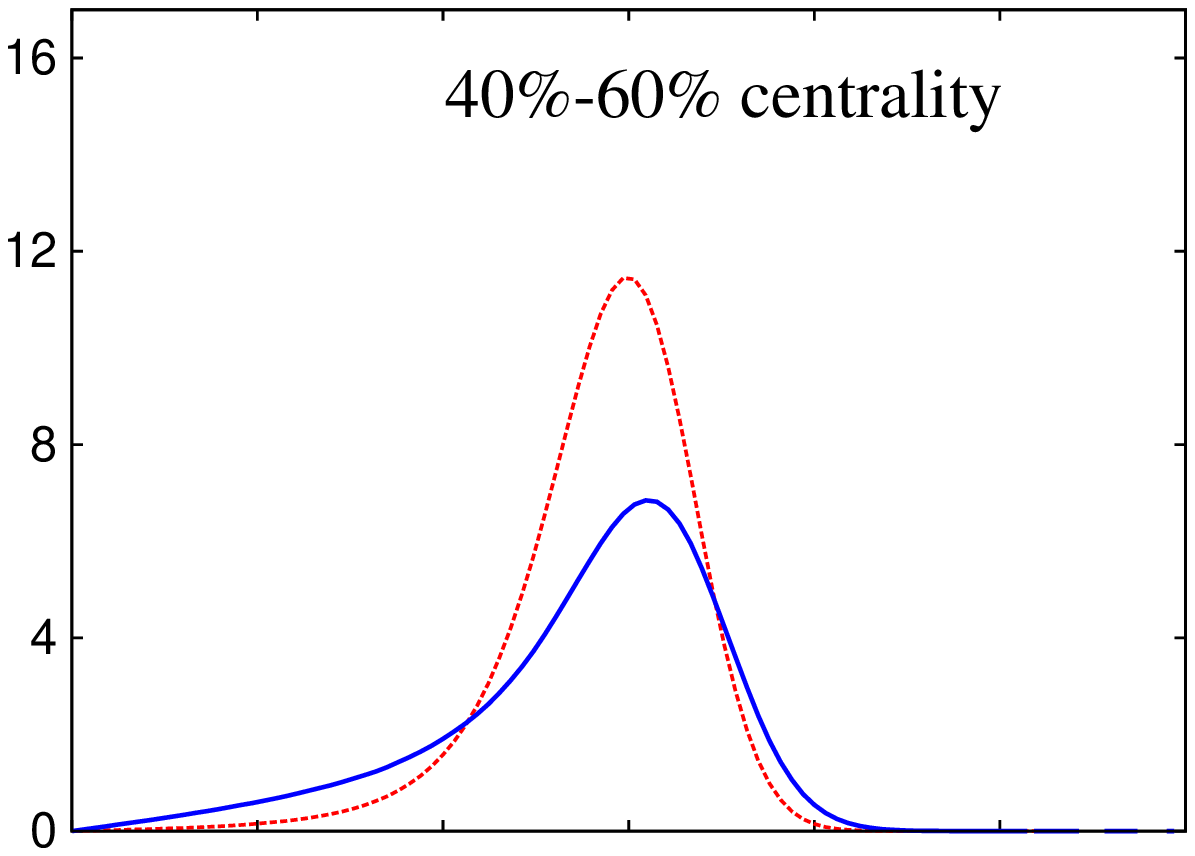}}
  \vskip -0.285cm
  \centerline{\hspace{0cm}
    \includegraphics[width=6.2cm, height=4.2cm]{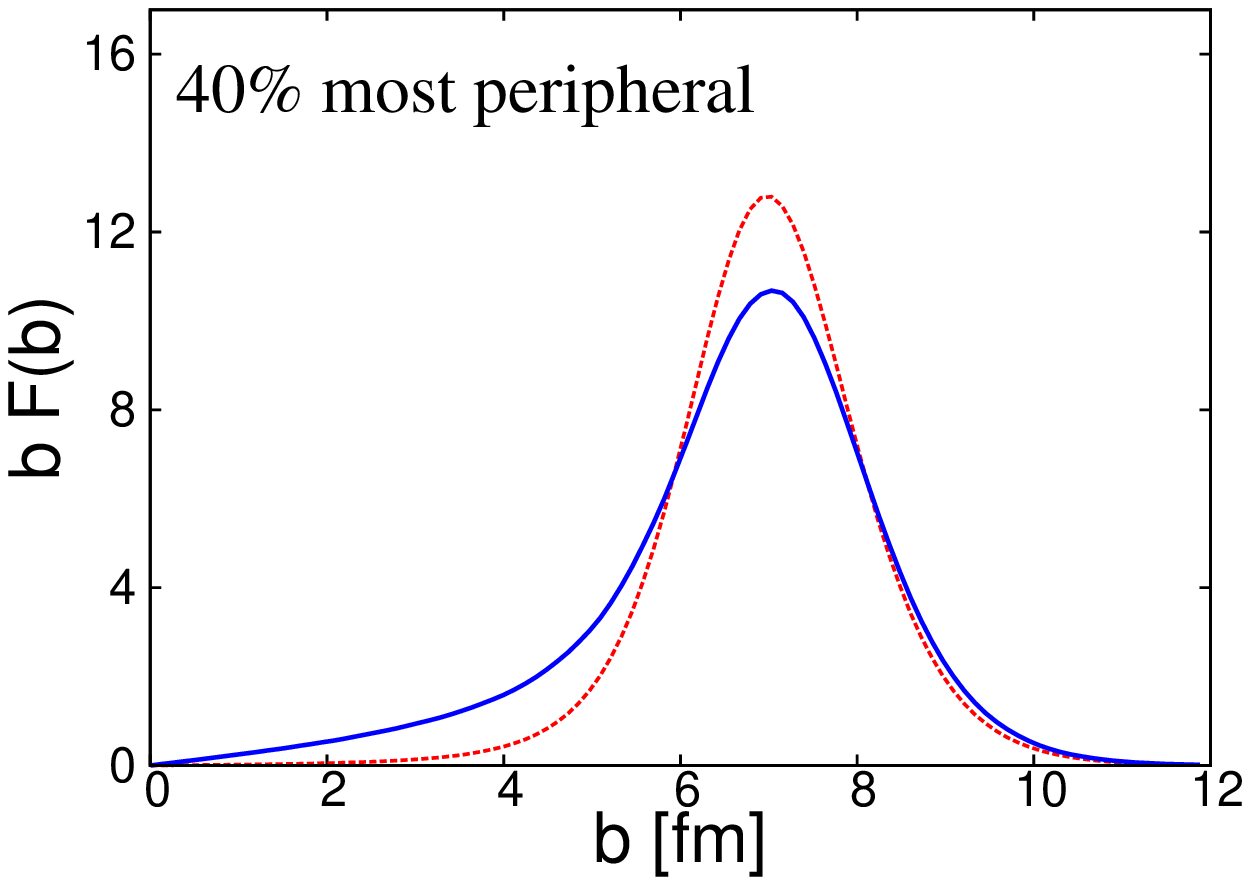}
    \hspace{-0.1cm}
    \includegraphics[width=5.8cm, height=4.2cm]{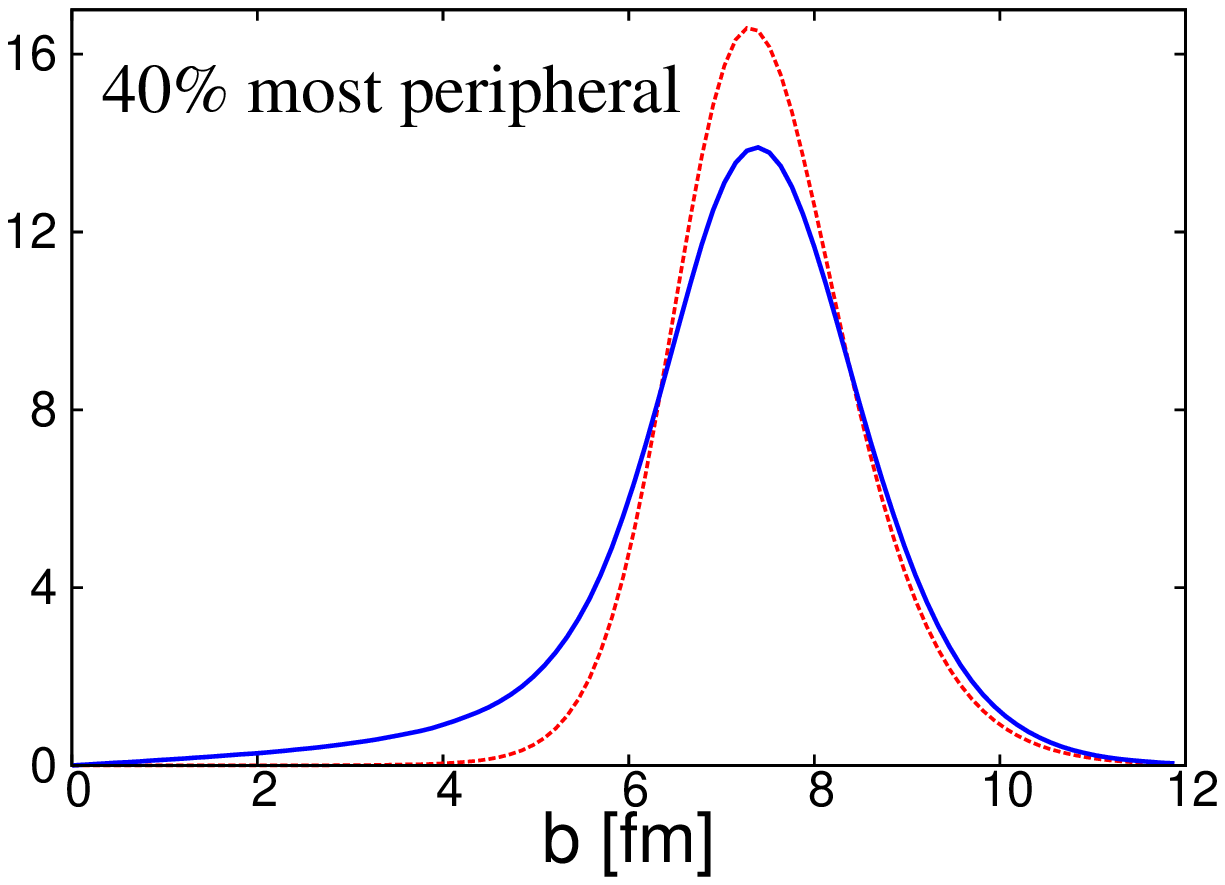}}
  \vskip -0.0cm
  \caption{The distribution over impact parameter, calculated with our MC, 
    of the different centrality classes 20\% most central (first row), 20\%-40\% 
    (second row), 40\%-60\% (third row), 40\% most peripheral (last row), both for 
    RHIC (left) and LHC (right) energies. Red: Glauber result; blue: Gribov - Glauber
    color fluctuations with $P(\sigma)$ distribution.
  }\label{figSIX}
\end{figure}

We also investigated the impact of fluctuations of the definition of centrality classes. We followed the 
experimental definition, in which the centrality is proportional to the fraction of total inelastic cross
section provided by a given type of  events. We can extract from the MC results of Fig. \ref{figONE} the 
probability $Q_N$ of having at least $N$ inelastic interactions, irrespective of the impact parameter $b$ 
(cf. Eq. (\ref{eq7.2a})):
\beq\label{qun}
Q_N\,=\,\frac{\sum^A_{M=N}\int d\Vec{b}\,P_M(b)}{\sum^A_{M=1}\int d\Vec{b}\,P_M(b)}\,,
\eeq
in such a way that $Q_{N=1}=1$ by definition. This allows to estimate the fraction 
of $\sigma^{hA}_{in}$ arising from a given interval in the number of wounded nucleons.
Then, one can choose a centrality class and select the interval
in number of wounded nucleons which contributes to that class. In Fig. \ref{figFIVE}, 
we have chosen the classes of the 20\% most central events by requiring it to provide 
20\% of the total inelastic cross section and, similarly, we have singled out the 
20\%-40\% and 40\%-60\% centrality classes, and the 40\% most peripheral 
events as the last class. We use the number of the wounded nucleons corresponding 
to (closer to) these cuts as limits in $N$ entering in Eq. (\ref{bfb}), for the 
calculation of  the curves in Fig. \ref{figSIX}. In Fig. \ref{figSIX} we show, for 
the selected classes,
% ======================================================================== Fig FIVE
%% =================================================================== End Fig FIVE
the distribution of events as a function of impact parameter by plotting
\beq\label{bfb}
b\,F(b)\,=\,b\,\sum^{N_{max}}_{N=N_{min}}\,P_N(b)\,,
\eeq
where $N_{min}$ and $N_{max}$ are the values singled out by the cuts described above
and shown in Fig. \ref{figFIVE}. The distribution of Fig. \ref{figSIX} was calculated 
both with the usual Glauber approach, \textit{i.e.} with a fixed $\sigma^{hN}_{tot}$, 
and with the inclusion of a fluctuating cross section according to $P_h(\sigma_{tot})$. 
For all the classes and both the RHIC and LHC energies, it can be seen that the fluctuating 
cross section tends to push the distribution toward larger values of the impact parameter.
% ======================================================================== Fig SIX
%% =================================================================== End Fig SIX

%------------------------------------------------------------------------------%
%------------------------------------------------------------------------------%
\section{Conclusions}
We have demonstrated that color fluctuations lead to a significant modification of the 
distribution over the number of nucleons involved in inelastic proton - nucleus collisions 
at collider energies. Study of the correlations between the soft central multiplicity and 
the rate of hard parton-parton interactions in the pA collisions at the LHC would provide 
a new avenue for investigating the three dimensional structure of proton. In particular 
such measurement will allow to test a conjecture that quark-gluon configurations in the 
proton containing large $x_p$ partons have a significantly smaller than average size.
%------------------------------------------------------------------------------%
%------------------------------------------------------------------------------%
\section{Ackowledgements}
We thank L.Frankfurt and V. Guzey for useful discussions during the preparation of the manuscript.
%------------------------------------------------------------------------------%
%------------------------------------------------------------------------------%

%------------------------------------------------------------------------------%
%------------------------------------------------------------------------------%
\end{document}